\documentclass[preprint]{aastex}

\begin{document}


\title{The Distribution of Maximum Relative Gravitational Torques
in Disk Galaxies}


\author{R. Buta\altaffilmark{1}, E. Laurikainen\altaffilmark{2}, and H. Salo\altaffilmark{2}} 



\altaffiltext{1}{Department of Physics and Astronomy, University of Alabama, Box 870324, Tuscaloosa, AL 35487 USA}
\altaffiltext{2}{Department of Physical Sciences, University of Oulu, Oulu,
Box 3000, Fin 90014, Finland}


\begin{abstract}
The maximum value of the ratio of the tangential force to the mean
background radial force is a useful quantitative measure of the
strength of nonaxisymmetric perturbations in disk galaxies.
Here we consider the distribution of this ratio, called $Q_g$,
for a statistically well-defined sample of 180 
spiral galaxies from the {\it Ohio State University Bright Galaxy Survey} and the 
{\it Two Micron All-Sky Survey}. 
$Q_g$ can be interpreted as the maximum gravitational torque 
per unit mass per unit square of the circular speed, and
is derived from gravitational
potentials inferred from near-infrared images under the assumptions of
a constant mass-to-light ratio and an exponential vertical density law.  In order 
to derive the most reliable maximum relative torques, orientation parameters
based on blue-light isophotes are used to deproject the galaxies, and
the more spherical shapes of bulges are taken into account using two-dimensional 
decompositions which allow for analytical fits to bulges, disks, and bars.
Also, vertical scaleheights $h_z$ are derived by scaling the radial scalelengths
$h_R$ from the two-dimensional decompositions allowing for the type dependence of 
$h_R/h_z$ indicated by optical and near-infrared studies of edge-on spiral 
galaxies. The impact of dark matter is assessed using a ``universal rotation
curve" parametrization, and is found to be relatively insignificant for our 
sample. In agreement with a previous study by Block et al. (2002),
the distribution of maximum relative gravitational torques is asymmetric towards large
values and shows a deficiency of low $Q_g$ galaxies. However, due to 
the above refinements, our distribution shows more low $Q_g$ galaxies
than Block et al. We also find a significant type-dependence in 
maximum relative gravitational torques, in the sense that $Q_g$ is lower 
on average in early-type spirals compared to late-type spirals.
The effect persists even when the sample is separated into bar-dominated
and spiral-dominated subsamples, and also when near-infrared
types are used as opposed to optical types.
\end{abstract}




\section{Introduction}

Nonaxisymmetric features are a pervasive and complex aspect of disk galaxies.
In normal, relatively non-interacting galaxies, these features are
in the forms of bars or spirals. It is well-known that the presence
of nonaxisymmetric structures in galaxy disks can impact the evolution
of morphology. For example, bars may drive spiral density waves
(Kormendy and Norman 1979), generate resonance rings of gas (Schwarz
1981; Buta \& Combes 1996), impact abundance gradients (Martin \& Roy 1994),
or induce gas inflow that may lead to bar destruction 
and bulge growth (Norman, Sellwood, \& Hasan 1996). A spiral
may trigger shocks, inducing star formation (Roberts, Roberts, \&
Shu 1975), or may rearrange stochastically-induced star-forming
regions into a more organized pattern (McCall 1986). It is clear
that nonaxisymmetric features, with their associated pattern
speeds and resonances, are extremely important to galactic evolution,
and understanding how these features develop is one of the principal
problems in galaxy formation and dynamics.

The source of much of the evolution caused by bars and spirals is
gravity torques due to tangential forces. Combes \& Sanders (1981;
see also Sanders \& Tubbs 1980)
suggested that these forces could provide a useful measure of the
strengths of nonaxisymmetric features such as bars, if the
potential could be determined. The idea is to derive the maximum
value of the ratio of the tangential force to the mean background
(or axisymmetric) radial force, which would give a single 
dimensionless number indicating the relative importance 
of nonaxisymmetry in the potential of a galaxy. This ratio,
which is physically the same as the maximum gravitational torque
per unit mass per unit square of the circular speed, will be
referred to in this paper as $Q_g$, while the method for
deriving $Q_g$ will be referred to as the gravitational torque 
method (or GTM).

The advent of routine near-infrared imaging of galaxies has made application
of the GTM more practical than ever.
Near-infrared images trace the stellar mass distribution of galaxies,
due to their emphasis on the older, dominant stellar populations.
Potentials can be derived from such images
using fast Fourier transform techniques in conjunction with assumptions
concerning the mass-to-light ratio and 
the vertical density distribution (e.g., Quillen, Frogel, and Gonz\'alez 1994, 
hereafter QFG). From this potential, the radial and tangential
components of the forces in the plane of the galaxy can be derived, and 
the Combes \& Sanders ratio can be estimated. Recent studies by Buta \&
Block (2001), Block et al. (2001), Laurikainen, Salo,
\& Rautiainen (2002), Laurikainen \& Salo (2002), 
and Block et al. (2002) have provided the first attempts 
to derive the maximum force ratios
for significant samples of galaxies. However, in these cases, the
samples were either ill-defined statistically, 
based entirely on relatively short exposure Two Micron All-Sky Survey (2MASS,
Skrutskie et al. 1997) near-infrared images,
or used deprojected images that did not allow for the typically
rounder shapes of bulges or the most reliable estimates of
vertical scaleheights.

There are good reasons for trying to derive the maximum force ratio
for a large, statistically well-defined sample of galaxies using a
refined version of the GTM.
Firstly, Sellwood (2000) has argued that we could evaluate scenarios of
bar formation in disk galaxies if we knew the observed distribution
of bar strengths. Various bar formation scenarios,
such as the natural ``bar instability" (Miller, Prendergast, \& Quirk
1970; Hohl 1971; Sellwood \& Wilkinson 1993 and other references therein)
or tidal bar formation (e.g., Noguchi 1996; Miwa \& Noguchi 1998), 
may predict different distributions of maximum relative 
bar torques, and an observed distribution may distinguish which mechanism
is most important. Secondly, recurrent bar formation due to accretion
of external gas would impact the distribution of maximum force ratios
(Bournaud \& Combes 2002). The idea is that bars can be the engines of
their own destruction in the presence of gas (see, for example,
Das et al. 2003), but may reform or
regenerate later if a galaxy accretes significant quantities of
external gas during a Hubble time that may cool the disk sufficiently (see
also Sellwood \& Moore 1999).
Thus, accretion can impact the ``duty cycle" of bars. This idea
was evaluated by Block et al. (2002) using an application of the GTM to the
Ohio State University Bright Galaxy Survey (OSUBGS, Eskridge et al.
2002). Block et al. concluded that the distribution of maximum
relative torques favored the idea that galaxies accrete enough
gas to double their mass in 10$^{10}$ years. 

In this paper, we re-examine the distribution of maximum relative torques
in spiral galaxies based on application of a much refined version of the 
GTM to basically the same OSUBGS sample as used by Block et al., supplemented
by a few larger galaxies with images from the 2MASS database. Our goal is 
to derive a reliable distribution of maximum relative bar and spiral 
torques in disk galaxies that can be compared with model predictions.
The refinements we use account for the shapes of bulges, improved estimates of
the galaxy orientation parameters, vertical scaleheights inferred
from type-dependent scalings of the radial scalelength, and a
statistical evaluation of the impact of dark matter. The $Q_g$
values we use are from Laurikainen et al. (2003). Only a few of the 
technical details connected with these values will be provided here,
and we refer the reader to Laurikainen et al. (2003) for a full
accounting of our application of the GTM. Our approach allows us to 
derive the most reliable maximum relative torques, and therefore the 
most accurate distribution of these torques.

\section{Properties of the Sample}

Our sample consists of 158 galaxies from the OSUBGS having inclinations
less than 65$^{\circ}$ and 22 2MASS galaxies having a similar
inclination limit but which were too large to be in the OSUBGS. 
The selection criteria for the OSUBGS are that the RC3 T index
is in the range 0$\leq T \leq$9 (S0/a to Sm), the total magnitude
$B_T$$\leq$12.0, the isophotal diameter $D_{25}\leq$6\rlap{.}$^{\prime}$5,
and the declination is in the range $-80^{\circ} < \delta < +50^{\circ}$
(Eskridge et al. 2002). Table 1 summarizes several of the mean properties
of the sample, based on data from RC3 (de Vaucouleurs et al. 1991).
Of the 180 galaxies, 177 have family 
classifications given in RC3. Table 1 shows that in the sample,
there are virtually equal numbers of galaxies classified
as SA, SAB, or SB.
Table 1 divides the averages according to this classification 
parameter. The table shows that mean parameters in the sample
are similar within these families. The mean Hubble type is
Sb-Sbc. Average colors, apparent angular size, radial
velocities, and distances are similar among the families.
There is an indication that, on average, the SA galaxies in the
sample are slightly more inclined than the SAB and SB galaxies.
Also, SA galaxies are slightly more luminous and larger
than SAB and SB galaxies. An inclination effect on the 
morphological recognition of bars is not unexpected and merely
highlights the difficulty of seeing bars which are weak and
viewed at high inclination. However, with bulge/disk
decomposition and deprojection, as well as near-IR imaging,
we can detect some of these lost bars.

\begin{figure}
\figurenum{1}
\plotone{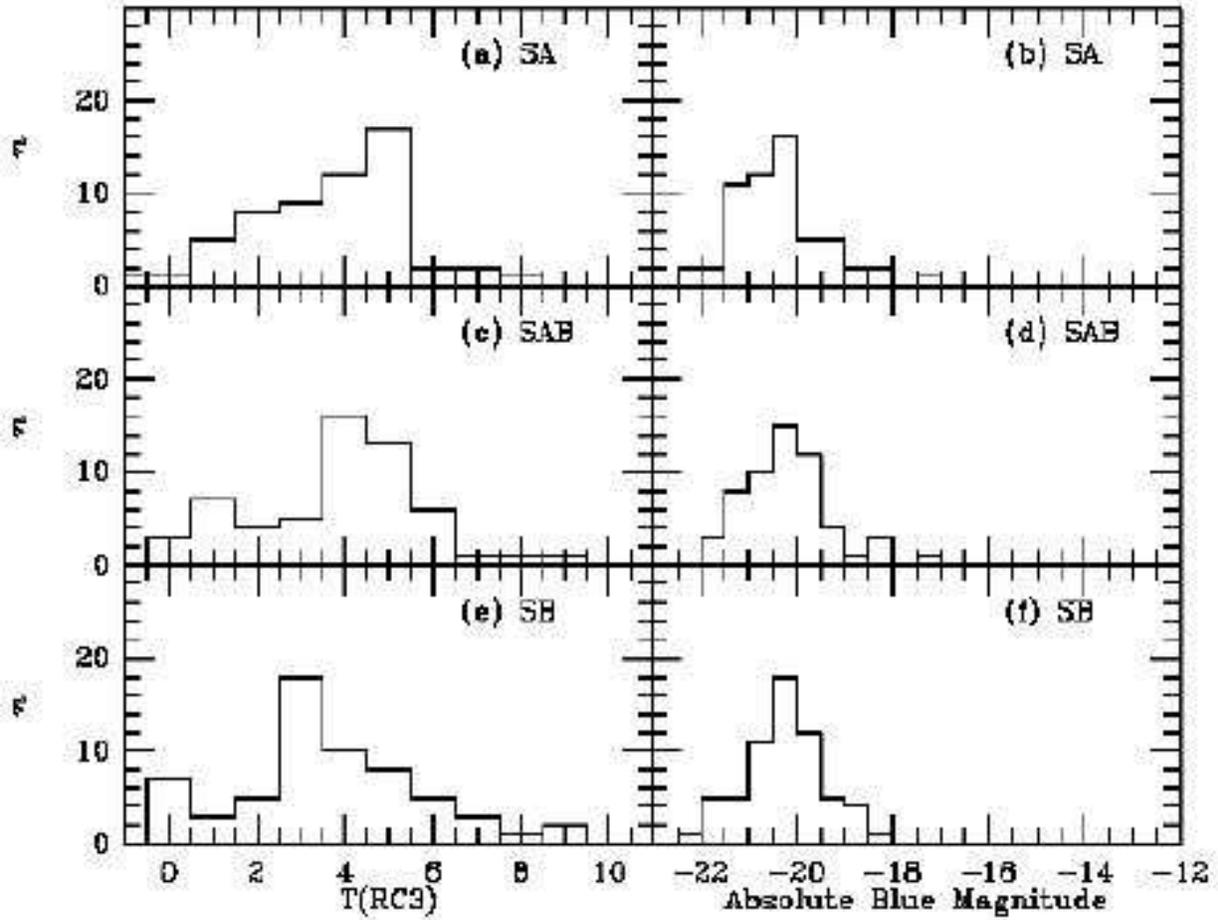}
\caption{Histograms of the number of sample galaxies, divided by RC3 family, versus
RC3 type index and absolute $B$-band magnitude, the latter 
based on RC3 magnitudes $B_T^o$, and on distances from Tully (1988) or the linear Virgocentric
flow model if not in that catalogue.
}
\label{histo1}
\end{figure}

\begin{figure}
\figurenum{2}
\plotone{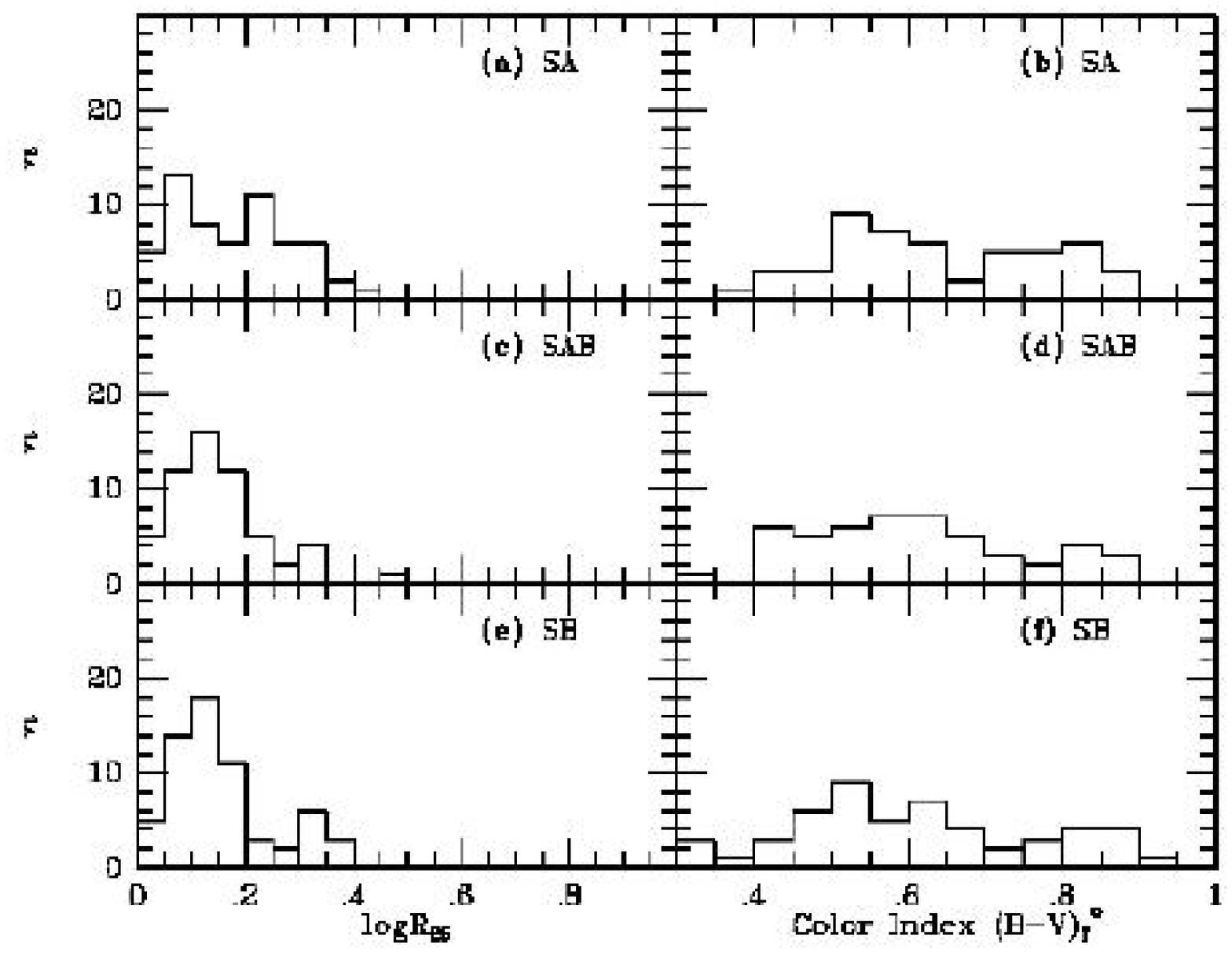}
\caption{Histograms of the number of sample galaxies, divided by RC3 family, versus
isophotal axis ratio $log R_{25}$ and total color index $(B-V)_T^o$, 
both parameters from RC3.
}
\label{histo2}
\end{figure}

Figures~\ref{histo1} and ~\ref{histo2} show the more detailed
distributions of SA, SAB, and SB galaxies in the sample
versus RC3 type, absolute blue magnitude $M_B^o$, the logarithm
of the isophotal axis ratio $log R_{25}$, and corrected color 
index $(B-V)_T^o$. Absolute magnitudes use $B_T^o$ from RC3 and
distances either from or on the scale of Tully (1988). 
Although the mean $T$ index is nearly the
same for the separate families, SB galaxies are asymmetrically distributed
towards early types while SA galaxies are asymmetrically distributed
towards later types. The distributions by absolute magnitude
show the higher luminosities of the SA galaxies compared to SAB
and SB galaxies. The distribution with $log R_{25}$ definitely emphasizes
lower inclinations for SB galaxies, while it is more uniform
for SA galaxies to the cutoff. Integrated colors are similarly
distributed over the three families.

\begin{figure}
\figurenum{3}
\plotone{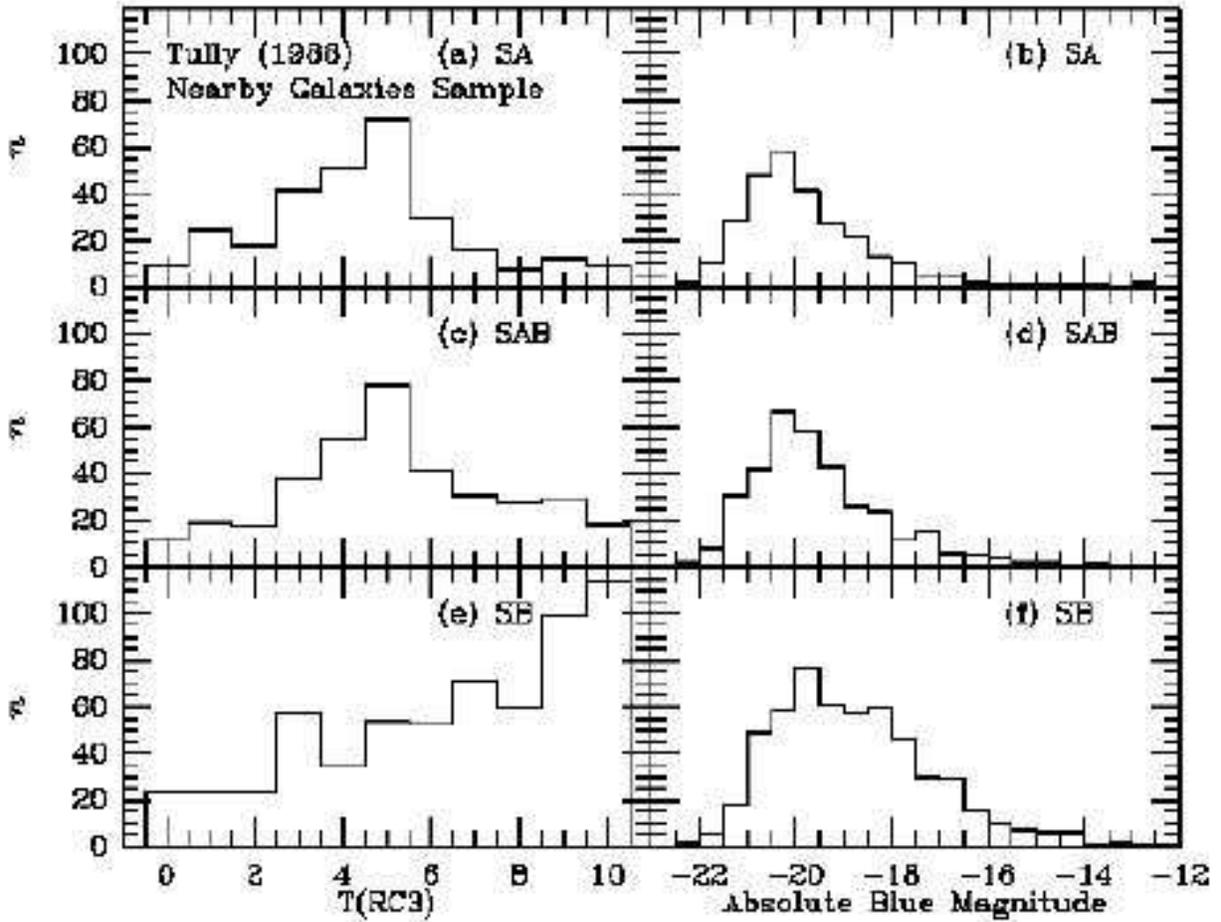}
\caption{
Histograms of the number of galaxies, divided by RC3 family, in the 
distance-limited sample of Tully (1988) versus
RC3 type index and absolute $B$-band magnitude, the latter 
based on RC3 magnitudes $B_T^o$.}
\label{histo3}
\end{figure}

\begin{figure}
\figurenum{4}
\plotone{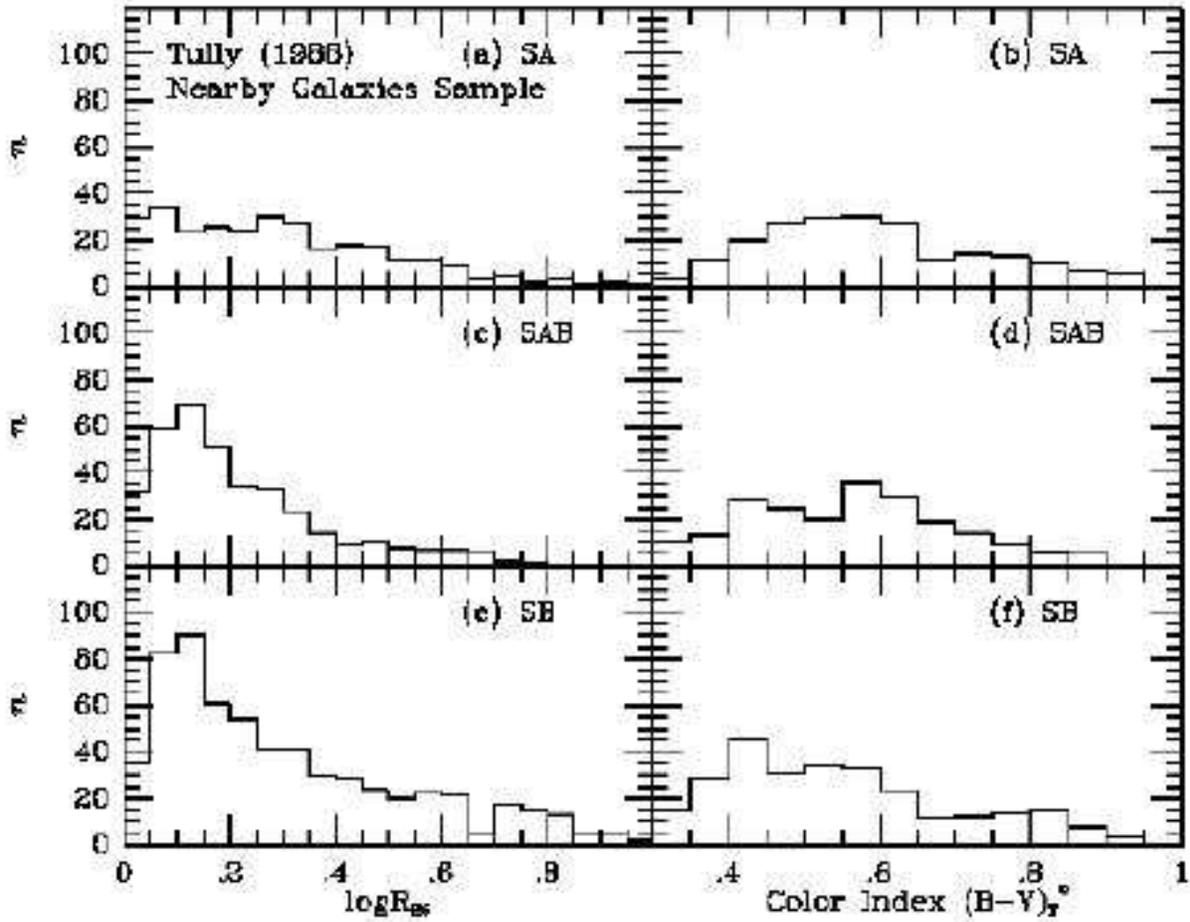}
\caption{
Histograms of the number of galaxies, divided by RC3 family, in the distance-limited
sample of Tully (1988) versus
isophotal axis ratio $log R_{25}$ and total color index $(B-V)_T^o$, 
both parameters from RC3.
}
\label{histo4}
\end{figure}

For comparison, Figures~\ref{histo3} and ~\ref{histo4}
show the same histograms for a distance-limited sample
of 1264 spirals\footnote{This number includes only those
Tully sample galaxies having RC3 data available.}
from the catalog of Tully (1988). Table 1
lists the mean parameters for the same sample. Our
magnitude- and diameter-limited OSU/2MASS sample emphasizes
earlier Hubble types and brighter absolute magnitudes
than the Tully catalog, the differences being most
extreme for SB galaxies. The distributions of color
and axis ratio, except for our inclination cutoff,
are similar to those for our sample galaxies. 
Thus, our sample is mainly biased
against late-type, low-luminosity barred spirals.
There is less bias in the SA and SAB subsamples
because these tend to have fewer late-type, low
luminosity examples. A critical issue is that it
appears that our sample is not necessarily biased
much against {\it nonbarred} spirals.

\section{Refinements to the GTM}

The basic assumptions in the GTM are: (1) the near-infrared light
distribution traces the mass, i.e., the mass-to-light ratio is constant; (2) the
vertical density distribution can be simply represented as, for example,
exponential with vertical scaleheight $h_z$; and (3) galaxies can be deprojected
as thin disks, after allowing for the shape of the bulge. As noted by
Buta \& Block (2001), the first assumption is probably valid for many galaxies
in the bar region, where maximum disks tend to be found (e.g.,
Freeman 1992). However, this is still an open question as
noted by Kranz, Slyz, \& Rix (2003), who used the amplitudes of 
modeled noncircular motions in five spirals to deduce that maximum disks
may be valid only if the maximum rotation velocity exceeds
200 km s$^{-1}$. In our sample, this would be the case only
for galaxies having $M_B$ $<$ $-$20.8 (Tully et al. 1998).
We address this issue further in section 8 using the ``universal
rotation curve" approach of Persic, Salucci, \& Stel (1996). 
Laurikainen \& Salo (2002) showed that the GTM is fairly 
insensitive to the form of the assumed vertical density 
distribution. 

\subsection{Polar versus Cartesian Grid}

The first refinement we use over Buta \& Block (2001) is a polar coordinate
grid as opposed to a Cartesian grid (Laurikainen \& Salo 2002). 
Buta \& Block used the QFG method of transforming near-IR
images into gravitational potentials, which operates on a 
two-dimensional image. This approach provides an
image of the potential, which can be used to derive a two-dimensional
map of the ratio of the tangential to the mean radial force. In such
a map, if a strong bar is present, four well-defined maxima or minima
are seen in the form of a ``butterfly pattern." Buta \& Block defined the
bar strength $Q_b$ to be the average of the absolute values of the
four maxima/minima.

Laurikainen, Salo, \& Rautiainen (2002) and Laurikainen \&
Salo (2002) used a polar grid approach as an alternative to QFG to
allow the application of the GTM to noisy and rather low resolution
2MASS images. Fourier components of the light distribution are computed
as a function of radius $R$ and azimuthal angle $\phi$, and these
Fourier light components are individually transformed into potential
components. The potential is then reconstructed analytically,
and the maximum force ratio, $Q_T = |F_T/F_{0R}|_{max}$, 
as a function of radius is computed. 

\subsection{Orientation Parameters}

In previous GTM studies such as those of Buta \& Block (2001) and Block et al. 
(2001, 2002), orientation parameters from RC3 were used
to deproject most of the galaxy images. However, these orientation 
parameters are in many cases based on photographic images and can
be manifestly improved with modern digital images. We have used 
the $B$-band images from the OSUBGS to fit ellipses to outer isophotes
and derive mean axis ratios and position angles for the outer
disks. In the future, these can also be improved upon using
two-dimensional velocity fields. The results of the ellipse fits,
as well as uncertainties, will be provided by Laurikainen et al.
(2003).

\subsection{Bulge Shapes}

Although the bulges of some barred galaxies might be as flat as the disk 
(Kormendy 1993), in many galaxies the bulge is a rounder component
than the disk. If this rounder shape is ignored when deprojecting
a galaxy, the bulge isophotes will be stretched into a bar-like
distortion (called ``deprojection stretch" by Buta \& Block 2001), 
leading to false torques. 
To deal with this problem we have used two-dimensional photometric 
decomposition, based on S\'ersic models (S\'ersic 1968) 
and allowing for seeing effects. The bulge and disk are described as in 
Mollenh\"off \& Heidt (2001), and in addition a bar component is added to 
the fit (Ferrer's bar with index n=2), which in some cases is essential 
for avoiding artificially large bulge models. The technique we used, as well
as the derived parameters, will be outlined in more detail by Salo, 
Laurikainen, \& Buta (2003). The decompositions allowed
us to remove the bulges, deproject the disks, and then add back the
bulges as spherical components. Thus, our analysis is not affected
seriously by bulge ``deprojection stretch."

\subsection{Vertical Scaleheight}

The computation of a potential from a near-infrared image requires
a value for the vertical scaleheight, which can be directly measured only 
for edge-on galaxies. Buta \& Block (2001) and Block et al. (2001) simply assumed that 
all galaxies had the same vertical exponential scaleheight as our Galaxy, 
$h_z$ = 325pc (Gilmore \& Reid 1983). However, this approach required knowledge of the distance
to each galaxy, which had to be based on radial velocities. Here we
follow Laurikainen, Salo, \& Rautiainen (2002) and derive $h_z$ (=0.5$z_0$, where $z_0$ equals the
isothermal scaleheight)
by scaling values from the radial exponential scalelength, $h_R$. As
shown by de Grijs (1998), the ratio $h_R$/$h_z$ depends on Hubble
type, being larger for later types compared to earlier types. Values
of $h_R$ were provided by our decompositions, and we used the following
scalings by type: $h_R/h_z$ = 4 for S0/a-Sa galaxies, 5 for 
Sab-Sbc galaxies, and 9 for Sc galaxies and later.

\section{The Maximum Relative Gravitational Torque}

We define the maximum relative gravitational torque, $Q_g$, to be the
maximum value of the ratio of the tangential force to
the mean radial force derived from a plot of $Q_T$ versus
$R$, based on a quadrant analysis.
In some cases, $Q_g$ is mostly measuring the maximum torque 
due to a bar, while in other cases $Q_g$ is clearly measuring
only spiral torques. In many cases, $Q_g$ will be measuring a
combination of bar and spiral torques, as shown by Buta, Block,
\& Knapen (2003), who developed a Fourier-based bar/spiral
separation technique. Thus, our analysis cannot provide
a true distribution of maximum relative bar torques $Q_b$.
For the evaluation of accretion models of spirals, Block
et al. (2002) noted that this is not a problem because
the models often also have spiral torques that contribute
to $Q_g$ estimates. 

\section{The Distribution of $Q_g$ Values}

\begin{figure}
\figurenum{5}
\plotone{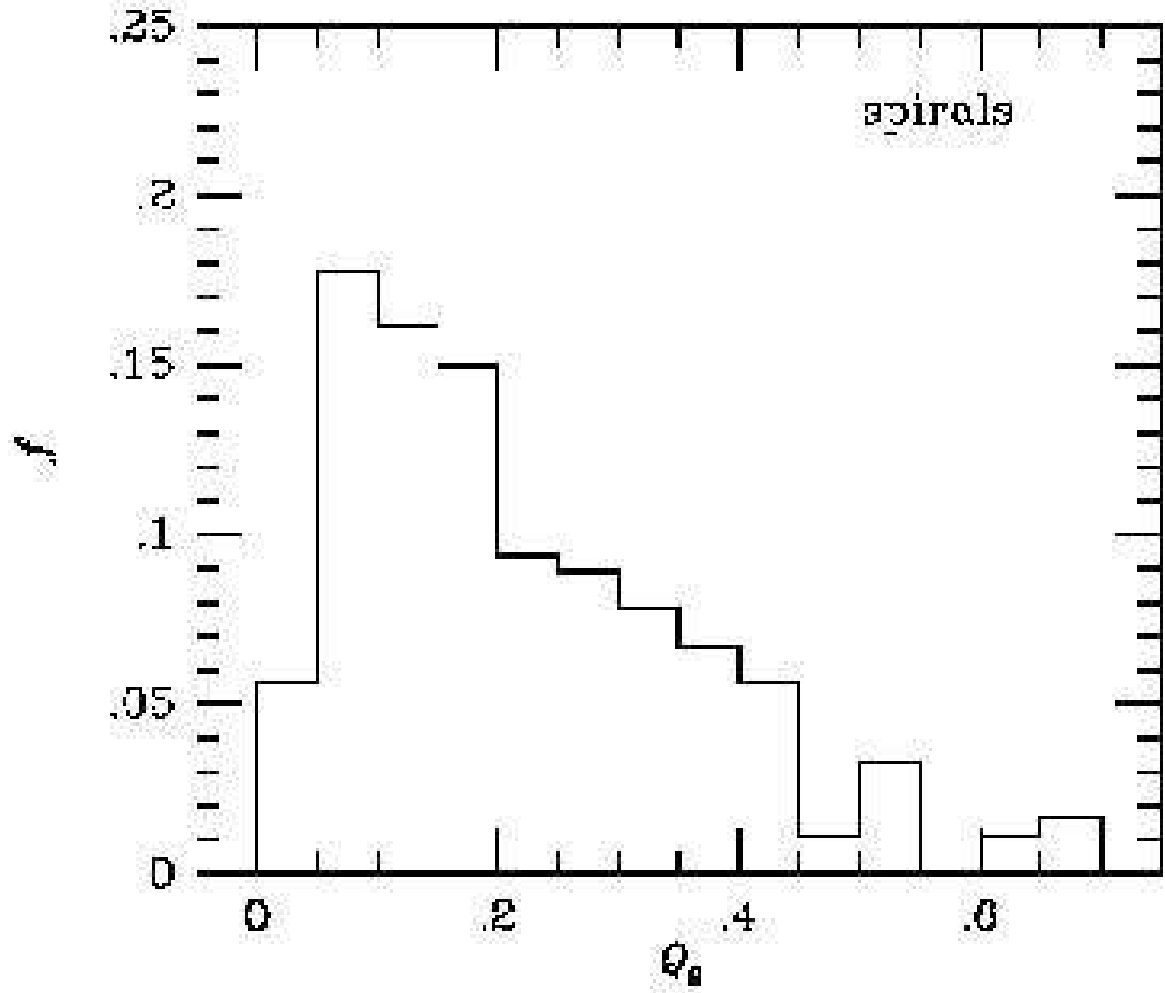}
\caption{
Relative frequency of maximum gravitational torques for 180
spiral galaxies.}
\label{freqs}
\end{figure}

\begin{figure}
\figurenum{6}
\plotone{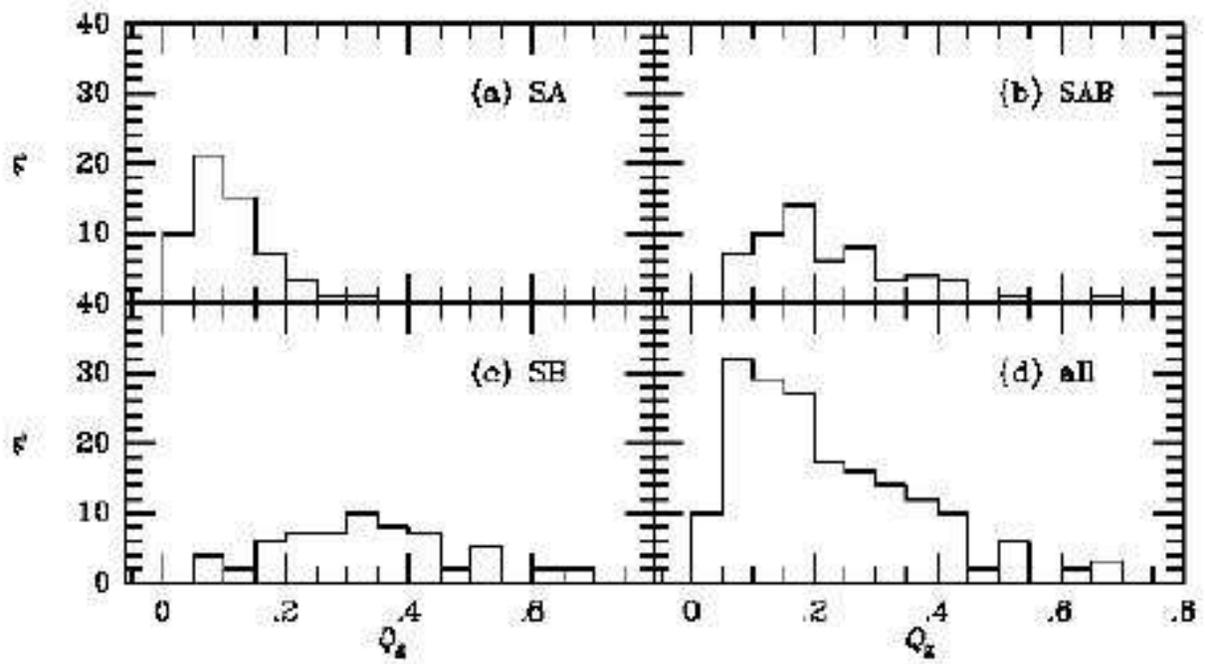}
\caption{
Histograms of the number of galaxies, divided by RC3 family and for the
full sample, versus the maximum relative gravitational torque $Q_g$ for the
OSUBGS/2MASS sample of 180 galaxies.
}
\label{distrib}
\end{figure}

Our main result is shown in Figure~\ref{freqs}, and is compiled
as counts $n$ and relative frequencies $f$ (=$n$/180) in Table 2. The distribution
of maximum relative gravitational torques is shown for the full sample of 
180 galaxies in comparison to the subsamples of SA, SAB, and SB galaxies
in Figure~\ref{distrib}.
The latter plots show again that there is indeed a correlation between
maximum torque and de Vaucouleurs family classification, but the
spread in $Q_g$ is very wide for SAB and SB galaxies. SA galaxies
appear to genuinely select the narrowest range of $Q_g$, while
SAB and SB galaxies include objects having $Q_g$ between
0.05 and 0.7. Thus, except for SA galaxies, the de Vaucouleurs
family classifications do not tell us much about real gravitational
bar torques except in an average sense. In Table 1, the mean values
of $Q_g$ by family are listed. The mean increases linearly from
SA to SB, with maximum relative gravitational
torques being 11\% for a typical SA galaxy, 
22\% for a typical SAB galaxy, and 33\% for a typical SB galaxy. 

Figure~\ref{freqs} shows an asymmetric distribution of maximum
relative gravitational torques, with a ``tail'' extending to
$Q_g$$\approx$0.7.  From the histograms in Figure~\ref{distrib}, 
it is clear that the primary peak in this
plot is due mainly to SA and SAB galaxies, while the extended tail
is due to SAB and SB galaxies. The average value of $Q_g$ for the
full sample is 0.222 with a standard deviation of 0.147. 

\section{Distribution Uncertainties and a Comparison with Block et al. (2002)}

As we have noted, a similar study of the distribution of maximum relative gravitational
torques in the OSUBGS sample was made by Block et al. (2002). They
selected 163 galaxies from the original sample of 198 having inclinations 
of 70$^{\circ}$ or less and not members of obviously interacting
systems. Vertical exponential scaleheights were derived from roughly 
estimated radial scalelengths (see below) as $h_z=h_R/12$. Most importantly,
no bulge/disk decompositions were made to allow for the likely rounder
shapes of bulges, and approximate orientation parameters from RC3
were used for the deprojections. Like us, however, Block et al.
derive $Q_g$ from graphs of $Q_T$ versus $R$.\footnote{Block et al. (2002)
use the term $Q_b$ for their parameter, but it is not derived in the
same manner as the $Q_b$ defined by Buta \& Block (2001). Instead,
it is the same as our definition of $Q_g$.} Thus, a comparison 
between our histogram of maximum relative torques and theirs is appropriate.

\begin{figure}
\figurenum{7}
\plotone{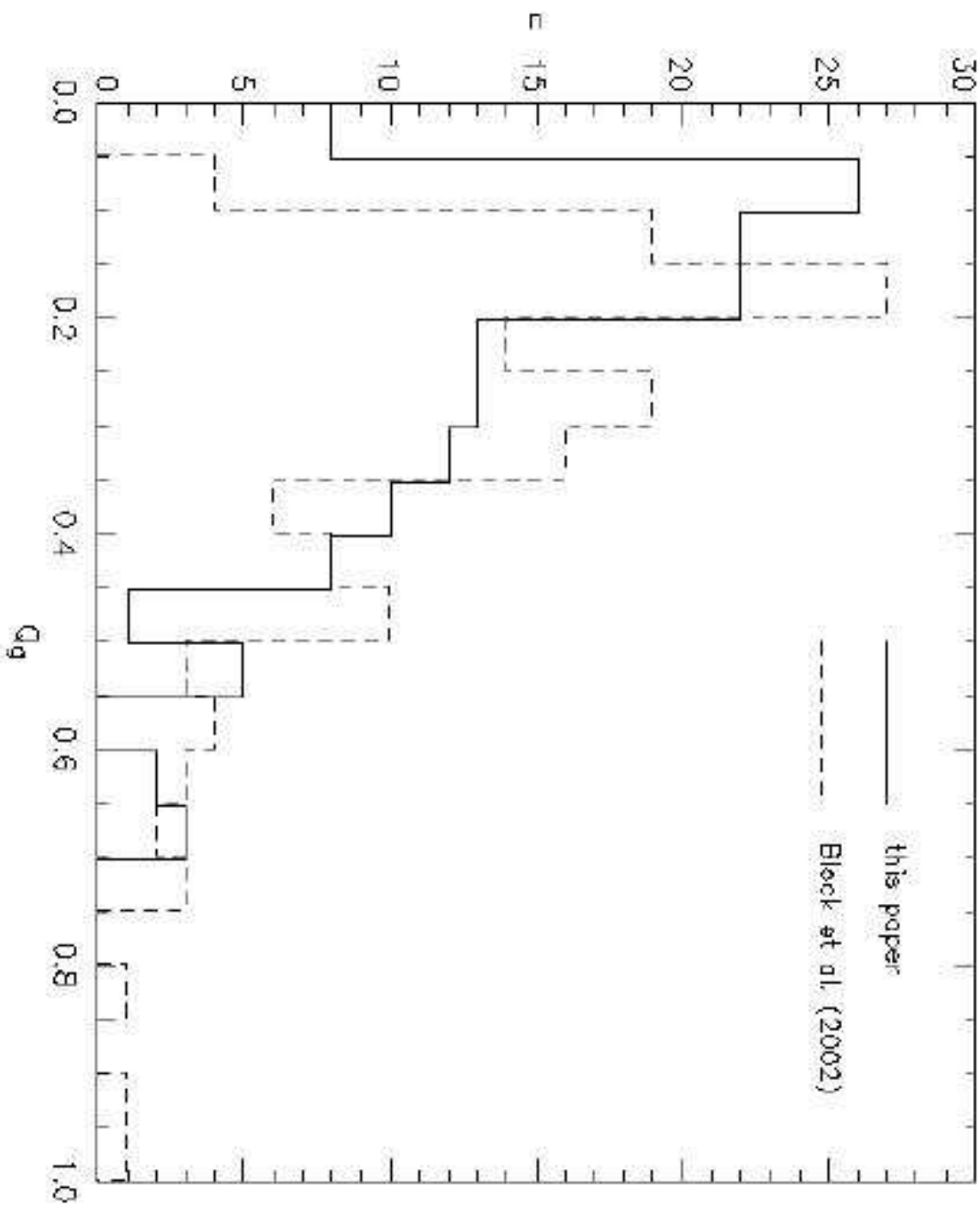}
\caption{
Comparison of the distribution of maximum relative gravitational
torques $Q_g$ for this paper (solid histogram) and Block et al. (2002)
(dashed histogram). The comparison sample includes 145 galaxies
from the OSUBGS only (see text).}
\label{compare}
\end{figure}

\begin{figure}
\figurenum{8}
\plotone{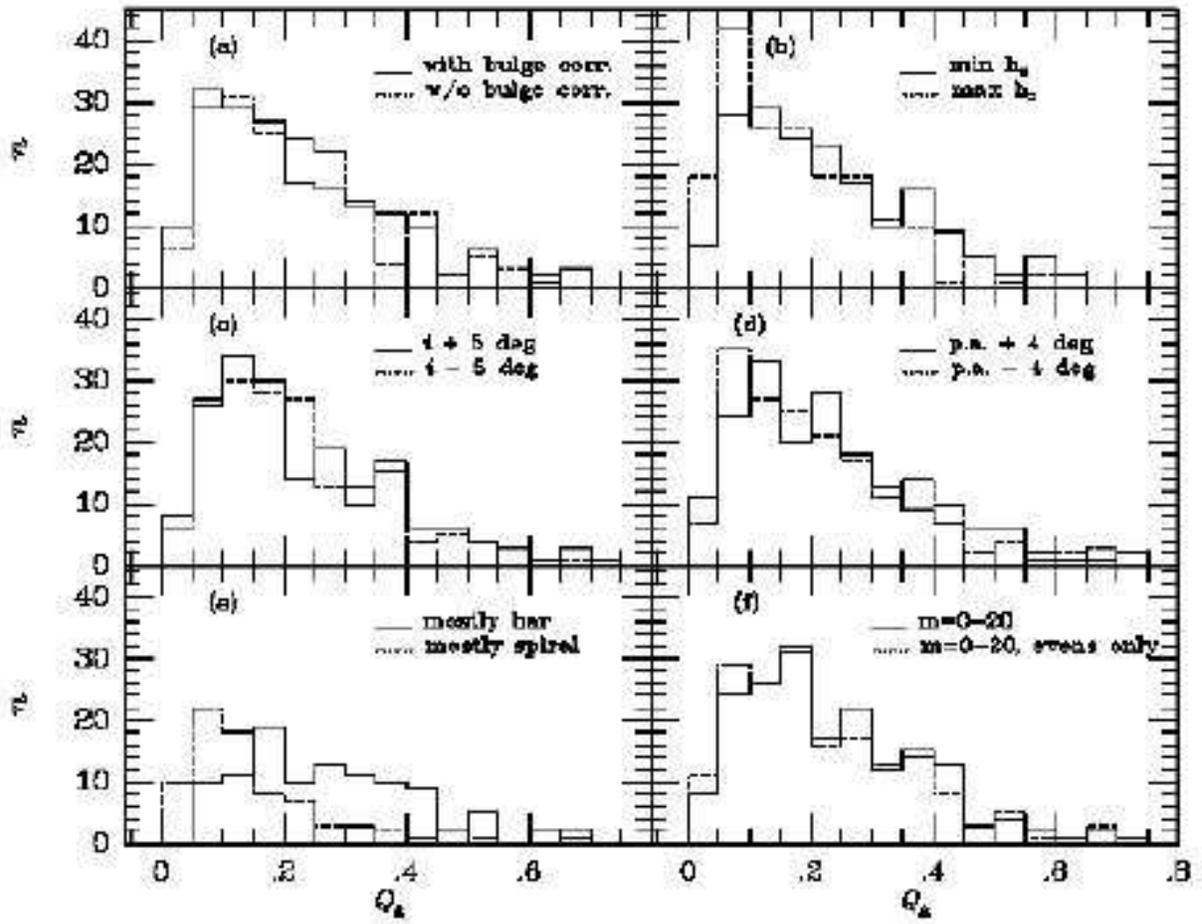}
\caption{
Histograms highlighting the impact of uncertainties due to
(a) bulge correction; (b) vertical scale height; (c) inclination;
(d) major axis position angle; (e) bar and spiral diagnostics;
and (f) number of Fourier terms on the distribution of maximum
relative gravitational torques.}
\label{errors}
\end{figure}

Figure~\ref{compare} compares the Block et al. distribution of
maximum gravitational torques with our distribution. 
The Block et al. histogram is not exactly the same as the one
published, but is based on a table kindly sent to us by. F. Combes.
It includes 159 galaxies where the measured $Q_g < 1$. In spite of
the similar numbers of objects, the Block et al. sample is missing
13 galaxies that are in our sample, and includes 18 galaxies missing
from our sample. The differences are in part due to our different
inclination cutoffs (65$^{\circ}$ in our analysis versus 70$^{\circ}$
used by Block et al.) as well as the different axis ratios used
to estimate inclinations (isophotal fits for our sample versus
RC3 $logR_{25}$ for Block et al.). To make the comparison fair,
we use only the 145 galaxies in common between our samples.
Although both histograms are similar in showing an asymmetric distribution,
our distribution shows more galaxies having low maximum relative
torques ($Q_g$ $\leq$ 0.15). The first two bins in the Block et al. 
histogram are extremely deficient in galaxies, a point used by them 
to argue that galaxies double their mass by accretion in 10$^{10}$ 
years. The reasons for the differences can be tied directly to a number
of causes, highlighted by the histograms in Figure~\ref{errors}.
Figure~\ref{errors}a shows that without the correction for bulge
shape, deprojection stretch can depopulate the first two bins.
However, the effect seems less important than might have been
expected given that our inclination cutoffs were high in both
cases. A more serious effect could be the assumed scaleheights,
as shown in Figure~\ref{errors}b. In this plot, we allow for the
scatter in $h_R/h_z$ from de Grijs (1998) and compute $Q_g$
for the minimum values of $h_R/h_z$ = 1, 3, and 5 (``max $h_z$" case)
and maximum values of 5, 7, and 12 (``min $h_z$" case) for types 
S0/a to Sa, Sab-Sbc, and Sc and later, respectively. The 
``max $h_z$" case clearly shows more low $Q_g$ values than the
``min $h_z$" case. Since Block et al. (2002) used $h_R/h_z$ = 12 for
all galaxies irrespective of Hubble type, their analysis favored
lower vertical scaleheights and larger values of $Q_g$ on average. 
Our use of bulge/disk decompositions and a type dependence 
to $h_R$/$h_z$ means that on average, our vertical scaleheights
are higher than those used by Block et al. (2002), and hence our
gravitational torques will be weaker. For a fairer comparison,
we have recomputed $Q_g$ for our deprojected images assuming
$h_z=h_R/12$. As expected, this depletes the first two bins but does not account 
for all the differences seen. The use of improved
orientation parameters could also contribute a little to the
differences.

Figures~\ref{errors}c and d show that uncertainties of $\pm$5$^{\circ}$
in inclination and $\pm$4$^{\circ}$ in major axis position
angle do not impact the observed distribution of gravitational 
torques too seriously. The number of Fourier terms to $m$=20
(Figure~\ref{errors}f) also has little impact.

Figure~\ref{errors}e shows the histograms for those galaxies
where $Q_g$ is clearly measuring a bar mostly and those 
where $Q_g$ is clearly measuring a spiral. The distinction
was made by examining the phase of the $m$=2 component
in the region of the maximum. If this phase was relatively
constant, then the $Q_T$ plot was concluded to be bar-dominated
at the radius of the $Q_g$ maximum. Otherwise, it was
concluded to be spiral-dominated. Both distributions
show a wide spread, although spirals are weaker on
average than bars.

Table 3 summarizes the uncertainties in individual estimates
of $Q_g$ due to inclination, position angle, and vertical
scaleheight. The table compiles the average deviation for
$i\pm$5$^{\circ}$, $\phi\pm$4$^{\circ}$, and the minimum
and maximum values of $h_z$, for three bins of inclination. 

\begin{figure}
\figurenum{9}
\plotone{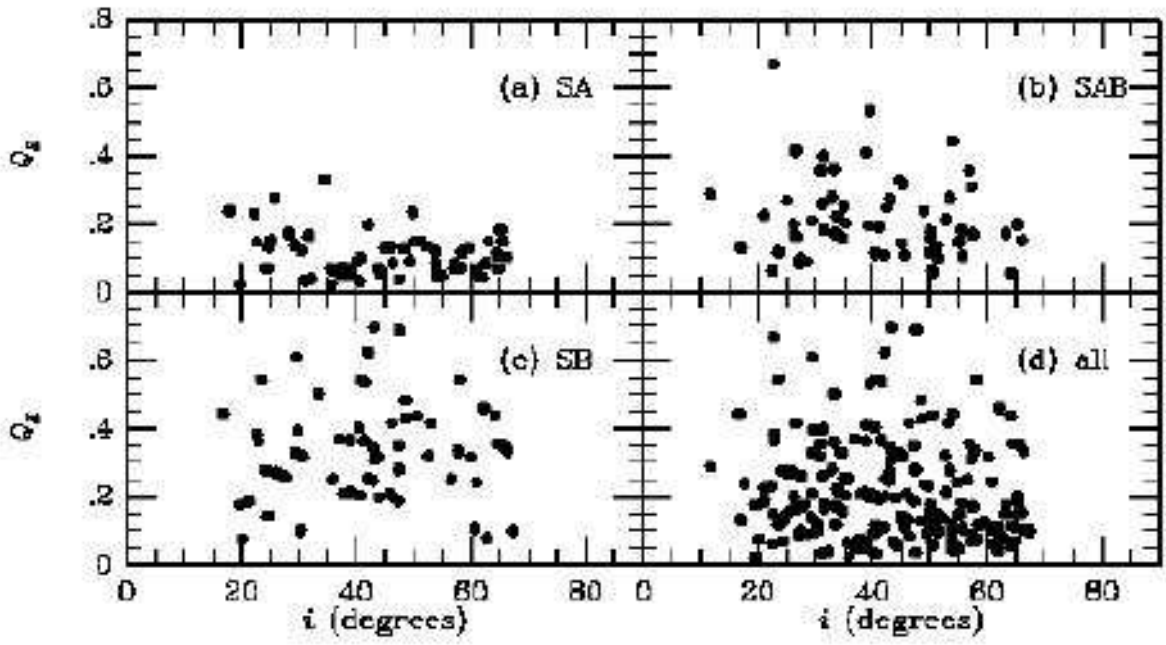}
\caption{
Plots of $Q_g$ versus inclination $i$ for the SA, SAB, SB, and
full samples.}
\label{incanal}
\end{figure}

In Table 4 and Figure~\ref{incanal}, we look for any systematic
effects due to inclination. Figure~\ref{incanal} shows plots
of $Q_g$ versus inclination $i$, where $i$ is computed using
either our mean ellipse-fit axis ratios for the OSUBGS sample,
or log$R_{25}$ for the 2MASS sample. We compute $i$ assuming
oblate spheroids and an intrinsic axis ratio $q_0$=0.2.
The figure shows no strong systematic effect with inclination.
This is verified in Table 4, where we compile the mean
$Q_g$ values for each sample in Figure~\ref{incanal}
divided around the median: 
45\rlap{.}$^{\circ}$6 for the SA sample,
40\rlap{.}$^{\circ}$7 for the SAB sample,
42\rlap{.}$^{\circ}$6 for the SB sample, and
42\rlap{.}$^{\circ}$7 for the full sample. Except for the
SAB sample, the high and low inclination samples
have the same means within the mean errors.

\begin{figure}
\figurenum{10}
\plotone{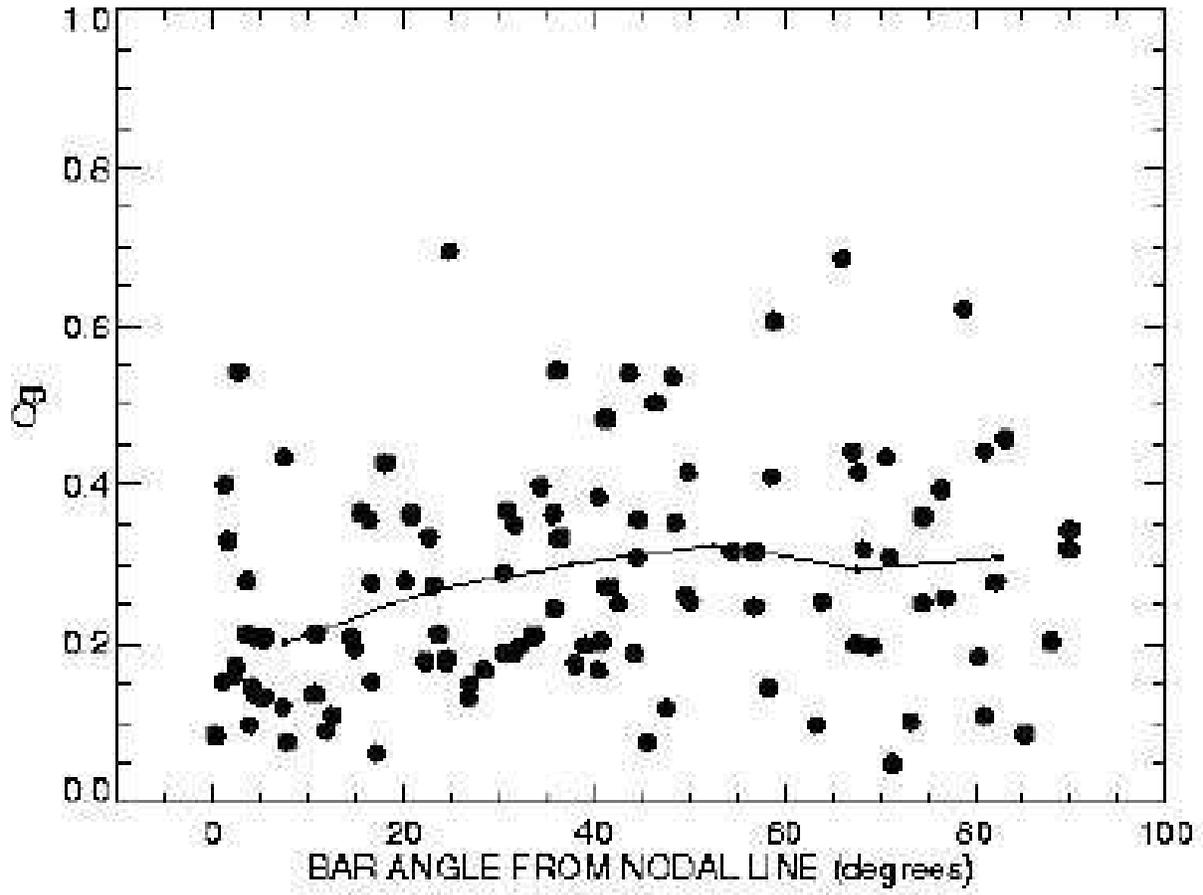}
\caption{
Plot of $Q_g$ versus relative projected bar position angle, 
$\phi_b$. The solid curve is the running mean of $Q_g$ in 15$^{\circ}$
bins.}
\label{barang}
\end{figure}

Another issue related to uncertainties is the impact of the
position angle of the bar relative to the line of nodes. Buta \& Block
(2001) showed that in a case like NGC 1300, where the bar is oriented
nearly along the line of nodes, the maximum torque is very sensitive
to the assumed inclination. The same would be true if the bar
is viewed end-on. We have investigated how important this might
be in our current sample. Figure~\ref{barang} shows a plot of
$Q_g$ versus relative bar position angle $\phi_{b}$. In this
plot, $\phi_b$ is determined from the phase of the $m$=2 
component of the potential at the radial location where $Q_T$
attains a maximum; the direction in the disk plane is then
projected to the sky plane. Analysis of Figure~\ref{barang}
indicates that there is indeed a bias in the sense that the
average bar strength is weaker for those systems where the
bar becomes ``thicker'' in deprojection. The averages are

$$for\ \phi_b\ <\ 30^{\circ}, <Q_g>\ =\ 0.223$$
$$for\ \phi_b\ >\ 30^{\circ}, <Q_g>\ =\ 0.303$$

\noindent
The solid line in the plot shows the running mean of $Q_g$
in 15$^{\circ}$ wide bins. The difference is statistically
significant, with the probability of having the same true mean
values being only 0.0035.

\begin{figure}
\figurenum{11}
\plotone{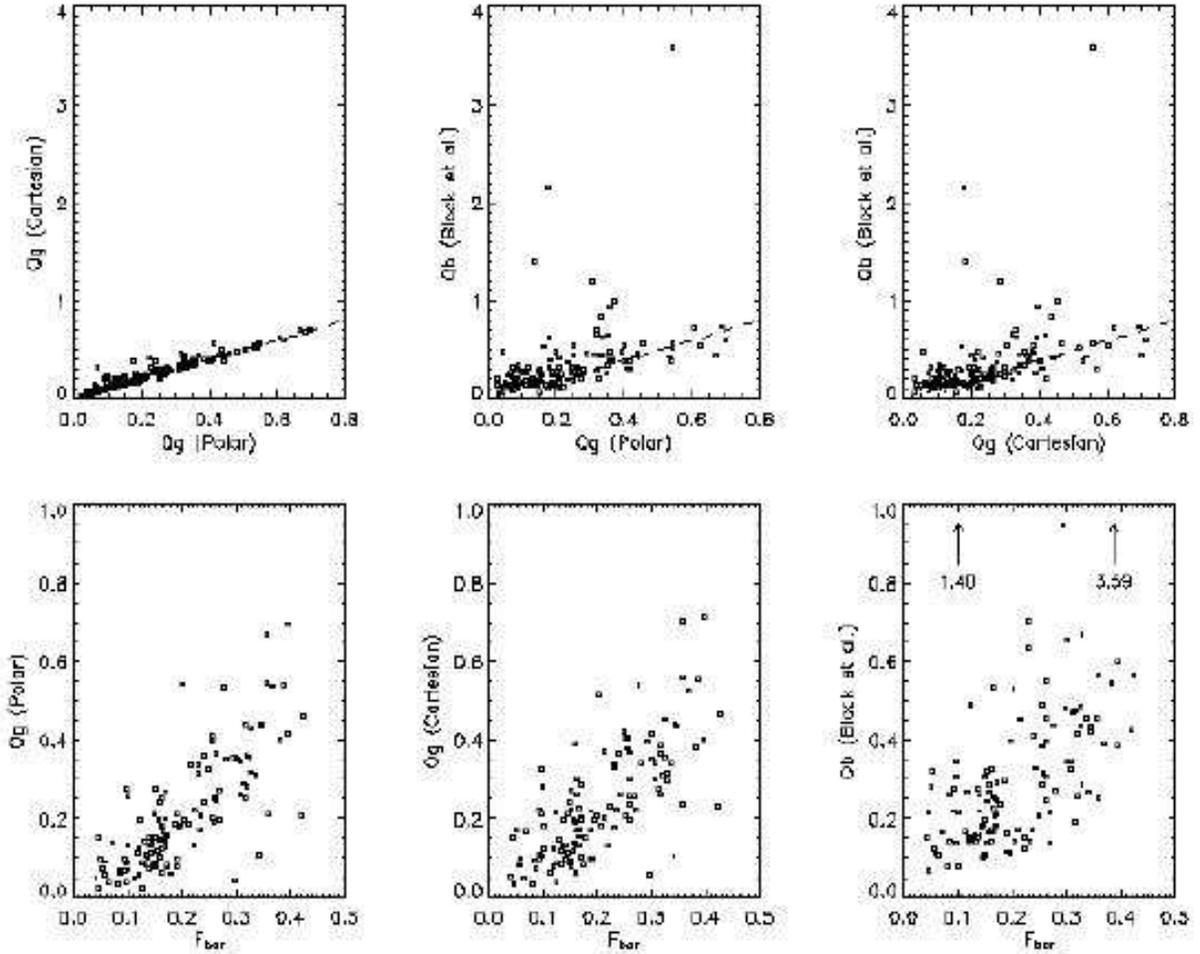}
\caption{
Top frames: Comparisons of $Q_g$ estimated using Cartesian and polar grid approaches
to estimating the gravitational potential. The upper left panel compares our
estimates from both approaches, while the upper middle and upper right panels
compare our values with the $Q_b$ estimates of Block et al. (2002).
Bottom frames: Plots of our estimates of $Q_g$ from polar and Cartesian
grid approaches and the $Q_b$ estimates Block et al. (2002) 
with the Whyte et al. (2002) bar strength parameter $f_{bar}$.}
\label{whyte}
\end{figure}

The referee has questioned whether our use of a polar grid
approach might cause lower values of $Q_g$ to be measured.
The idea is that smoothing with a polar grid might reduce
the strength of the perturbation, increasing the number of
low $Q_g$ values. We have checked this
by recomputing our $Q_g$ values using a Cartesian approach
with a 128$\times$128 grid resolution (covering the whole
galaxies usually, but not necessarily the whole image).
The radial profiles $Q_T(R)$ were constructed separately
for four image quadrants, and the mean of these profiles 
was computed. The Cartesian $Q_g$ was then taken from the
peak of the Cartesian $Q_T(R)$ profile, limited to the radial
range around the force maximum found by the polar method.
This was done to insure that the Cartesian $Q_g$ corresponds
to the bar region, and does not refer to some spurious
force maximum in the outer parts of the images. Figure ~\ref{whyte}
(upper panels) shows the results of the comparison. We find very good agreement 
between our $Q_g$ estimates from the Cartesian and polar grid approaches. 
However, comparison of the same numbers with the Block et al.
(2002) values is poorer, as shown by the upper middle and
upper right panels of Figure~\ref{whyte}.

The upper left panel of Figure~\ref{whyte} does show that
some Cartesian $Q_g$ values are noticeably larger than the
polar grid values. However, as discussed in Laurikainen \& Salo (2002), 
the Cartesian method can lead to large
spurious force values in the noisy outer parts of images, sometimes
leading to an overestimate of $Q_g$ if the results are automatically
collected, without careful inspection of the force profiles.
This might account for several very large values of $Q_b$ estimated
by Block et al. (2002), seen in the upper panels of Figure~\ref{whyte}.
Mainly for this reason, we chose the polar grid force evaluation
as our standard procedure. The Cartesian method is useful 
as a check of the polar method results.

As a further check on how our methods affect the histogram
of maximum relative torques, we have analyzed more closely three highly-inclined
galaxies in our sample, NGC 3166, 3338, and 3675, trying to duplicate
the methods used by Block et al.: (1) use the RC3 position
angle and inclination to deproject the galaxies; (2) no
correction for the shape of the bulge; (3) radial scalelength
derived from $logD_{25}$ in RC3 assuming all the galaxies
follow the Freeman (1970) law, with $h_z$=$h_R$/12; and (4) using a Cartesian
transformation for the potential. The results 
are $Q_g$ = 0.26, 0.16, and 0.15, respectively, compared
with the values of 0.27, 0.14, and 0.15 actually derived
by Block et al. Thus, mimicking the Block et al. treatment
with our codes yields values that fully agree with those
obtained by Block et al. In contrast, our refined approach
gives values of $Q_g$= 0.11, 0.08, and 0.08 for the same
galaxies. The reason for the low $Q_g$ values we get compared to
theirs is due to our refinements, and not a serious
difference in our codes. 

The idea that galaxies might accrete significant quantities
of external gas during a Hubble time is certainly intriguing. 
Our revised histogram (with its extended tail of large
$Q_g$ values) still supports this idea, but may favor
an accretion rate between the two cases discussed by Block
et al. (2002): the no accretion idea and a rate which
doubles the mass in 10$^{10}$ years.
As shown in this work, the bulge correction, improvements in the 
orientation parameters, and the larger vertical scaleheights we use
considerably increase the number of galaxies with low maximum
relative torques. 

In spite of the differences with Block et al.,  we still find 
a deficiency of galaxies in the lowest torque bin, $Q_g$ $\leq$ 0.05.
Truly axisymmetric galaxies appear to be rare in the OSUBGS and 
2MASS samples, although we note that because $Q_g$ cannot be
negative, noise could also deplete the first bin to some extent.

\section{Comparison with the $f_{bar}$ Parameter}

Whyte et al. (2002) have used the OSUBGS to compute
bar strength using an isophotal analysis. They derived
a bar strength parameter, $f_{bar}$, based on the minimum 
$H$-band isophotal axis ratio, $(b/a)_{bar}$, in the bar region 
estimated from a moment analysis involving a series of 
cuts through an image in surface brightness (Abraham \& 
Merrifield 2000). The parameter $f_{bar}$ is convenient
because it scales the bar strength to the range 0.0 to 1.0,
and also because it stretches the range corresponding to the
important small $(b/a)_{bar}$ values. Block et al. (2002)
used the Whyte et al. results to support their findings
of few nonbarred galaxies in the OSU database, and thus
their conclusions concerning the accretion rate in galaxies.

The lower panels of
Figure~\ref{whyte} show comparisons between our $Q_g$ 
values (both polar and Cartesian) and
$f_{bar}$ and $Q_b$(Block et al.) and $f_{bar}$. The most
striking difference is how well $f_{bar}$ correlates
with our values of $Q_g$, showing that the shape of the
bar does correspond well to the strength of the gravity
field.  This was
also shown by Laurikainen, Salo, \& Rautiainen (2002) for their
2MASS sample. In contrast, the comparison between $f_{bar}$
and $Q_b$(Block et al.) shows a noticeably larger scatter.

In spite of the good agreement between $f_{bar}$ and our
$Q_g$ values, $f_{bar}$ is by no means a suitable replacement for
$Q_g$. $f_{bar}$ is probably determined by the 
self-consistent response of the bar to the gravitational
field that maintains it, and thus it measures the force
in an indirect fashion. $Q_g$, on the other hand, estimates
this field directly from the luminosity distribution.

\section{The Impact of Dark Matter}

Ideally, the way to assess the impact of dark matter on
a torque indicator such as $Q_g$ would be to compare an
observed rotation curve with a rotation curve predicted
from an azimuthally-averaged light profile, preferentially
a near-infrared profile corrected for color effects due
to a radial stellar population change (e.g., Bell \& de 
Jong 2001). Then the signature of the dark component 
would be how much the observed and predicted rotation
curves disagree, especially in the outer parts of the
galaxies. However, it is impractical for us
to carry out such a comparison for our full sample in
a homogeneous way. Thus, we have used a more statistical 
approach.

Our estimates for halo corrections are based on the extensive
analysis of rotation curves and light profiles by
Persic, Salucci, \& Stel (1996, hereafter PSS). In this paper
the dark halo rotation curves are described by the
isothermal sphere law, with a smooth transition to constant
core density

$${V_h}^2(x) = {V_{\infty}}^2  \frac{x^2}{x^2+a^2},   \eqno{1}$$

\noindent
where $x = R/R_{opt}$ is the radius normalized to the optical
radius, a fiducial
reference radius enclosing 83\% of the total blue luminosity.\footnote{For this
radius we have actually used $D_{25}$/2, which is specifically valid only
for a Freeman disk. The error committed for those galaxies that may not
be Freeman disks is not serious given the approximate nature of these
estimates.}
The parameter $a$ is the halo core radius, also in units of $R_{opt}$.
PSS (see especially their erratum) give, based on their 
sample of 1100 optical and radio rotation curves,

$$a=  1.5 (L/L_*)^{0.2} \eqno{2}$$ 

\noindent
and

$$\frac{\rm dark\ mass}{\rm visible\ mass} = 0.4 (L/L_*)^{-0.9} x^3
\frac{1+1.5^2 (L/L_*)^{0.4}}{x^2+1.5^2 (L/L_*)^{0.4}} \eqno{3}$$

\noindent
where $L_*=10^{10.4} L_{\odot}$ in the $B$-band.
Near the optical radius we may estimate

$$\frac{\rm dark\ mass}{\rm visible\ mass} \approx {V_h}^2/{V_d}^2\eqno{4}$$

\noindent
where $V_d$ includes the rotation velocity due to the disk plus bulge.

Eqs. (1)-(3) now define $V_h$ at all radii, as a function
of $L/L_*$, and the value of $V_d (x)$ at some value near $R=R_{opt}$.
Once $V_h(R)$ is known, the $Q_T(R)$ profiles computed under
a constant $M/L$ assumption are modified to

$$Q^{hc}_T(R) = \frac{Q_T(R) F_d(R)}{F_d(R)+F_h(R)} \eqno{5}$$

\noindent
where $F_d(R) = {V_d(R)}^2/R$ and $F_h(R) = {V_h(R)}^2/R$ are
the radial forces due to visible and dark masses, respectively,
and the superscript ``hc'' means ``halo-corrected''.
If the measurements extend to $R=R_{opt}$, then $V_d(x=1)$ has been
used, while in the case $R_{max}<R_{opt}$, then $V_d(x=R_{max}/R_{opt}$)
was used for fitting $V_h$. Values of $R_{opt}$ were taken from RC3,
and the $B$-band luminosities $L$ were calculated from $B$-magnitudes
and Galactic extinctions given in NED and distances from Tully (1988).

\begin{figure}
\figurenum{12}
\plotone{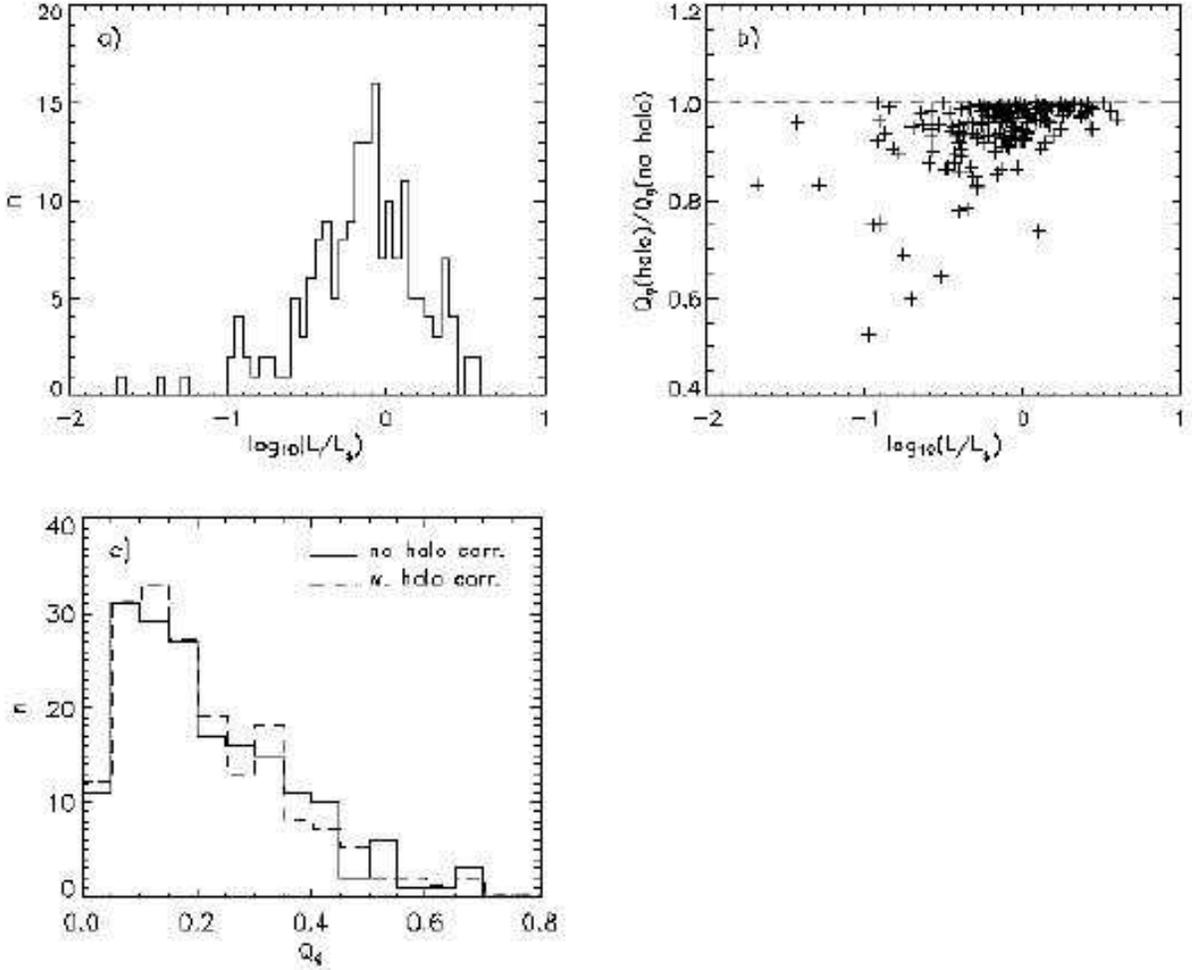}
\caption{
(a) Plot of the distribution of $L/L_*$ for
the sample galaxies, peaking near $L/L_*=1$.
(b) Plot of the distribution of
$Q^{hc}_g/Q_g$ (with/without halo correction), as a function
of $L/L_*$, indicating how the correction gets more important for less
luminous galaxies with more dominant halo components. The deviating
point at $L/L_* \approx 1.3$ is NGC 7213 (see text).
(c) Plot of the distribution of $Q_g$ with and without
halo correction. The similarity of the histograms shows
that dark matter has only a small impact on our results.}
\label{allhalo}
\end{figure}

Figure~\ref{allhalo}a shows the distribution of $L/L_*$ for
our sample of 180 galaxies. The distribution peaks near
$L/L_*$ $\approx$1. Figure~\ref{allhalo}b
shows the distribution of $Q^{hc}_g/Q_g$ as a function
of $L/L_*$, indicating how the correction gets more important for less
luminous galaxies with more dominant halo components. The deviating
point at $L/L_* \approx 1.3$ is NGC 7213, for which $Q_g$ is
practically zero and obtained near $R_{opt}$ ($Q_g$ changes from
0.023 to 0.017). Finally, Figure~\ref{allhalo}c
shows the distribution of $Q_g$ with and without
halo correction. The average value of $Q_g$ with the correction
is 0.209 compared with 0.222 without the correction, indicating 
only a marginal (6\%) reduction.

Altogether, the effect of dark halos appears to be weak for the sample,
which as we have shown is dominated by fairly luminous systems 
for which PSS models imply halos with rather large core radii and relatively
small mass within $R_{opt}$. Therefore, the contribution to $Q_T(R)$
is small in the inner parts of the galaxy where maximum $Q_g$'s
are typically obtained, at least for bars.
For spiral forces alone the effect would be more prominent.

A potential problem with the fits described above for low luminosity
galaxies is that in many cases the measurements
probably do not reach far enough, in terms of disk scalengths, to
yield reliable outer rotation curves (truncation of the disk overestimates
the disk radial force and thus the rotation velocities). For $Q_g$
measurements this is not a problem, as noted by Laurikainen \&
Salo (2002). However, the above procedure uses outer $V_d$'s to
estimate $V_h$'s, which therefore might in some cases be
overestimated. Indeed, strange, strongly rising rotation
curves follow for some of the less luminous galaxies when the above
procedure is applied (although they are rising already before
inclusion of the halo). Nevertheless, since this error in all cases
overestimates the reduction of $Q_g$ due to the inclusion of a halo, 
it is not important for the present purpose.

\section{Type Dependence of Maximum Relative Gravitational Torques}

Because the bulge is usually more significant in early-type galaxies,
we might expect that maximum relative gravitational torques would be
diluted somewhat compared to later-type galaxies. This is
because the bulge can be a significant contributor to the
mean axisymmetric radial force in the bar regions of early-type
spirals. Block et al. (2001) searched for this effect in their
combined sample of 75 galaxies but did not detect a measurable
type dependence. They argued that the bulge
dilution at early types could be partly offset by the shorter bars
found at later types (e.g., Elmegreen \& Elmegreen 1985).

Laurikainen, Salo, \& Rautiainen (2002) also searched for a
type dependence in $Q_g$ in a 2MASS sample of 43 barred galaxies, half
of which have AGN. In their sample,
19 galaxies have types Sa-Sb and 21 galaxies have types Sbc and
later. These authors derived $<Q_g>$ = 0.25$\pm$0.03 for the
early types and 0.38$\pm$0.05 for the later types, suggesting
a possible difference.

Our refined treatment of bulges and our larger sample
compared to these previous studies allows us to re-evaluate
this possible effect more reliably. As we have noted, we allowed for the
more spherical shapes of bulges using two-dimensional photometric
decompositions that took into account, where necessary, the
contributions of bars. We also treated bulges as spherical
in their potentials, such that the forces in the plain
are properly estimated. In Buta \& Block (2001) and Block et al. (2001),
bulges were assumed to be as flat as disks, which overestimated
their radial forces in the plane.

\begin{figure}
\figurenum{13}
\plotone{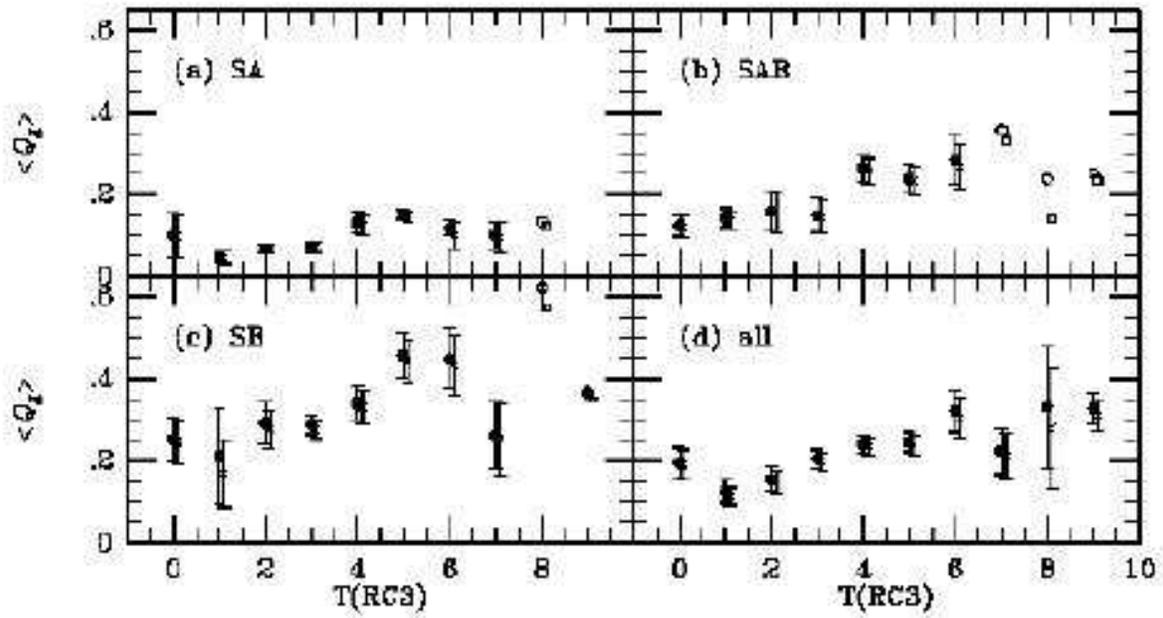}
\caption{
Plots of mean maximum relative torque versus RC3 type index. 
Error bars are mean errors. The filled circles show the means for no dark
halo correction, while the open circles indicate points based on only
one galaxy. The crosses show the means with a dark halo correction
and are offset by 0.1 in $T$ for clarity. Open squares indicate
halo-corrected values based on only one galaxy.}
\label{bytype}
\end{figure}

Figure~\ref{bytype} shows the correlation of $<Q_g>$ with RC3
revised Hubble type in our present sample. The filled circles
show the averages with no dark halo correction, while the crosses
show the averages with a halo correction. Table 5 also summarizes
the numerical values for no halo correction. This plot does appear to detect a type-dependence
in our measured maximum relative gravitational torques. For early-type
spirals ($T$=0-3, or S0/a-Sb), $<Q_g>$ = 0.177$\pm$0.014, while
for late-type spirals ($T$=4-9, or Sbc-Sm), $<Q_g>$ = 0.258$\pm$0.015.
A halo correction reduces these means only slightly, to 0.169
for S0/a-Sb and 0.247 for Sbc-Sm.
The difference between early and late-type spirals
appears to be significant. As shown in Figure~\ref{bytype}
and Table 4, the 
effect persists even when the sample is divided by de Vaucouleurs 
family, and has the same trend in the sense that early-types have
lower average $Q_g$. This suggests that
early-type spirals do indeed have diluted maximum relative
gravitational torques, an effect which must contribute to the 
observed scatter of $Q_g$ among the three de Vaucouleurs 
families. 

In interpreting this result, the first question one might ask is
how reliable the bulge decompositions are. Since we used a sophisticated
two-dimensional decomposition allowing for a bulge, a disk, and a bar
in the fit, we believe the decompositions are as good as we will be able
to make them. The referee argues that
bulge subtraction is delicate and not unique, and that if the
bulge participates in the bar instability (as in the box/peanut
shape), then its impact may not be reliably treated. This is a
valid concern. However, 
Laurikainen \& Salo (2002) have tested a radius-dependent scaleheight 
that simulates a peanut-shaped distribution in the sense that the vertical
scaleheight increases towards the outer parts of the bar by an amount similar
to that observed in real galaxies. This was found to affect $Q_g$ estimates
by only about 5\%. 

\begin{figure}
\figurenum{14}
\plotone{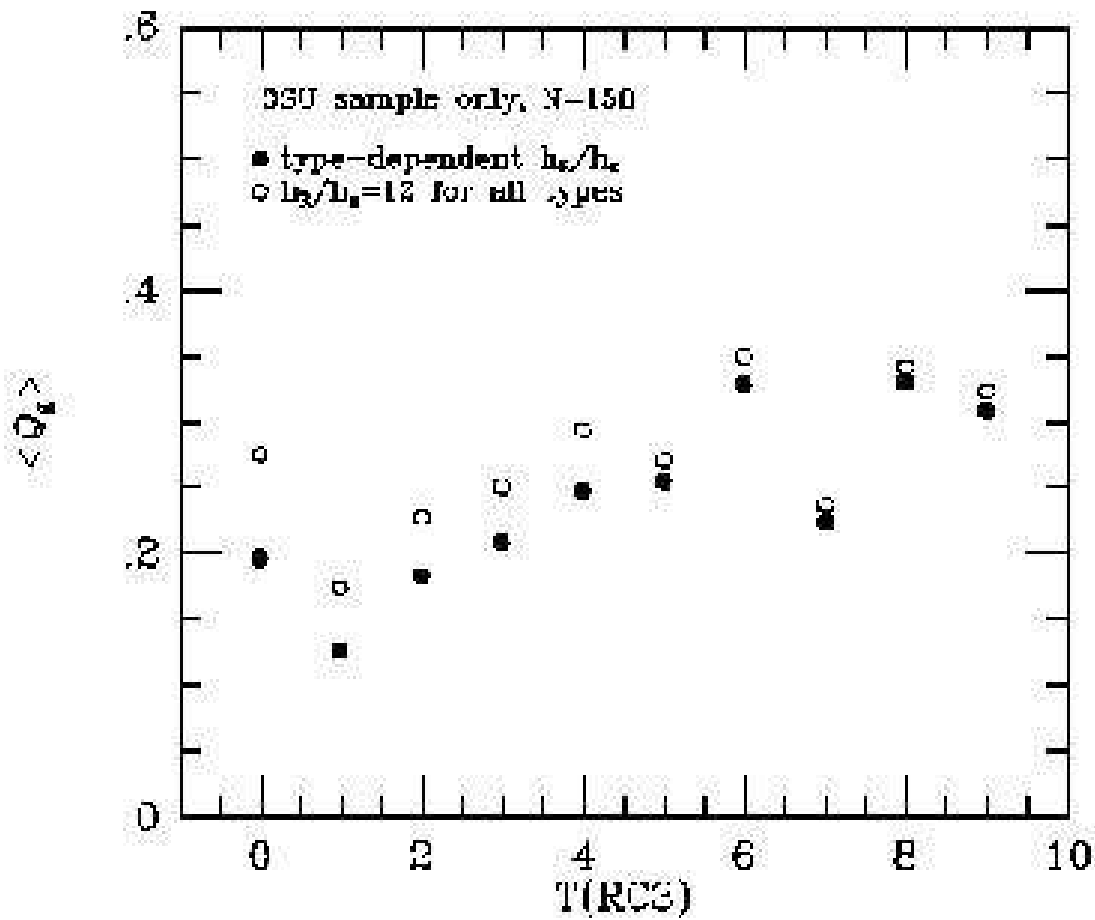}
\caption{
Plots of the mean maximum relative torque versus RC3 type index
for our full sample using the type-dependent ratio $h_R/h_z$ from
de Grijs (1998) and a type-independent ratio, $h_R/h_z$ = 12,
used by Block et al. (2002). Only OSUBGS galaxies are in these
samples.}
\label{bytype12}
\end{figure}

Another important question is how our assumptions concerning the
vertical scaleheight contribute to the observed type
dependence. Our estimates of $Q_g$ have
utilized the findings of de Grijs (1998) to infer $h_z$
from $h_R$, assigning larger values of $h_z$ to early-types
compared to late-types. If we assume instead that $h_z$
= $h_R$/12 for all types, we get the results shown in
Figure~\ref{bytype12}. Our assumption of a type dependence
to $h_R/h_z$ does indeed enhance the measured type-dependence
in $Q_g$. However, the assumption of a constant value
of $h_R/h_z$ is inconsistent with studies of edge-on
galaxies and favors our approach.

\begin{figure}
\figurenum{15}
\plotone{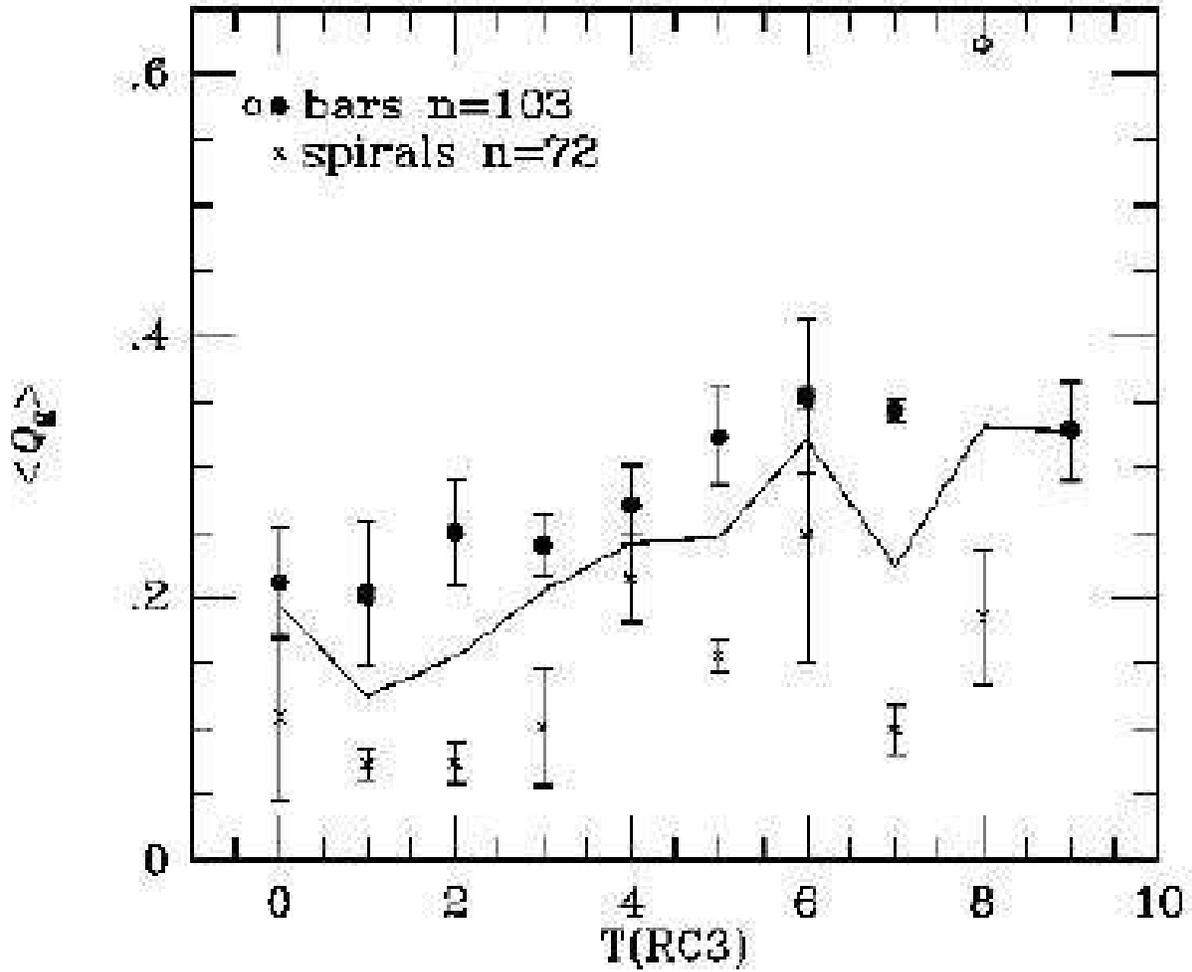}
\caption{
A plot of $<Q_g>$ versus RC3 type index separated according to
whether the radius of the $Q_T$ maximum occurs in the bar-dominated
region (filled circles) or the spiral-dominated region (crosses).
The open circle is based on only one galaxy.
The plot demonstrates that both spirals and bars have relatively
weaker torques in early-type spirals as compared to late-type
spirals. The solid curve shows the means from Table 5.}
\label{splitype}
\end{figure}

Figure~\ref{bytype} shows that $<Q_g>$ is type-dependent,
but it does not prove unequivocably that this means
bars are relatively weaker in early-type spirals than in
late type spirals. This is because $Q_g$ is also affected
by spiral arm torques. To try and approximately separate
the two phenomena, we use the bar/spiral discriminations
from Figure~\ref{errors}e and discussed in section 6. 
If we compute $<Q_g>$
as a function of type for these subsamples separately, we 
get the results in Figure~\ref{splitype}. Surprisingly,
it appears that both bars and spirals are relatively weaker
in early-types as compared to late-types. For bars especially,
the type dependence is remarkably well-defined.

\begin{figure}
\figurenum{16}
\plotone{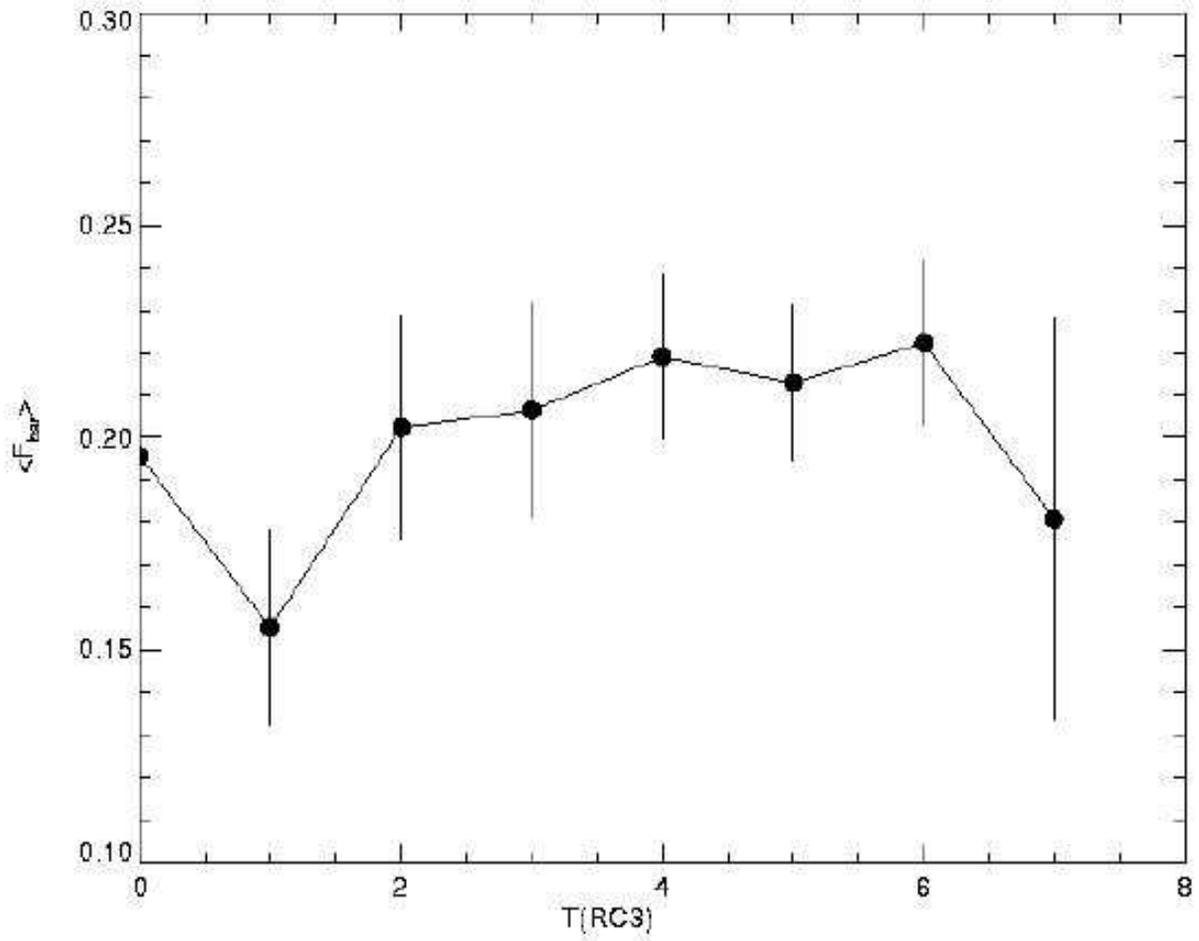}
\caption{
Plot of the average bar strength parameter, $f_{bar}$ from Whyte
et al. (2002) versus RC3 type index.}
\label{whyte2}
\end{figure}

A type-dependence in bar strength is also found in the Whyte et al.
(2002) analysis, although it is smaller than found for
$Q_g$. Figure~\ref{whyte2} shows $<f_{bar}>$
vs RC3 type index $T$. Just as for $Q_g$, early-type spirals
have lower average $f_{bar}$ than late-types. For 49
S0/a-Sb galaxies in the Whyte et al. sample, $<f_{bar}>$ = 0.190 $\pm$ 0.013,
while for 76 Sbc and later galaxies, $<f_{bar}>$ = 0.213 $\pm$
0.011. The effect is marginal but is still in the same
sense as found for $Q_g$.

Note that on the basis of theoretical models, one might
expect early-type galaxy bars to have stronger maximum
torques simply because the bars are longer than those
in later types (Elmegreen \& Elmegreen 1985). Apparently,
bulge dilution is a more dominant effect, so that late-type
galaxy bars are stronger in a relative sense. Note that this result 
refers mainly to Sbc-Sc galaxies as late-types, as our sample
has few galaxies of types Scd and later. This is a result
of our sample biases. A distance-limited sample would
provide more reliable results for the very late-type
spirals.

\section{Correlations with Near-Infrared Morphology}

Eskridge et al. (2002) used the $H$-band images in the
OSUBGS to estimate near-IR classifications of galaxies
within the revised Hubble framework of de Vaucouleurs
(1959) and Sandage and Bedke (1994). These classifications
include the family (SAB or SB and plane S for nonbarred
galaxies), and the stage from S0 to Sm. We converted the
$H$-band stages, estimated as if the images were blue
light images, to the RC3 numerical $T$ index scale.
Eskridge et al. (2002) note that the apparently increased
bulge-to-disk ratio and the greater degree of smoothness
of structure biases near-IR classifications towards
earlier types on average. For galaxies where these
effects changed the type from a spiral classification
to S0 or SB0, we have used the index $T$=$-$2.

Table 6 summarizes the mean values by stage and
family from the near-IR classifications. As noted
by Eskridge et al. (2000), near-IR classifications
from the OSU sample show twice as many strongly-barred
(SB) types as in the optical. However, Table 6 shows
that the Eskridge et al. SAB and SB classifications have
slightly lower $<Q_g>$ than the corresponding
RC3 families. RC3 SB galaxies in our sample have $<Q_g>$ = 0.331
$\pm$ 0.019 (m.e.), while Eskridge et al. SB galaxies
in our sample have $<Q_g>$ = 0.290 $\pm$ 0.015. The
likely reason for this difference is that near-IR images 
not only make weak bars more evident, but also
make stronger bars more obvious. Thus, near-IR
imaging does not necessarily change the rankings of bars
much. There is no new category for a $B$-band
SB spiral to be placed into even though it looks
stronger in the near-IR. However, a $B$-band
SAB spiral can be placed into the SB category if
it looks stronger in the near-IR. Since
the real rankings are not changed much, the mean
$Q_g$ for the near-IR families is decreased 
because of inclusion of weaker bars.

\begin{figure}
\figurenum{17}
\plotone{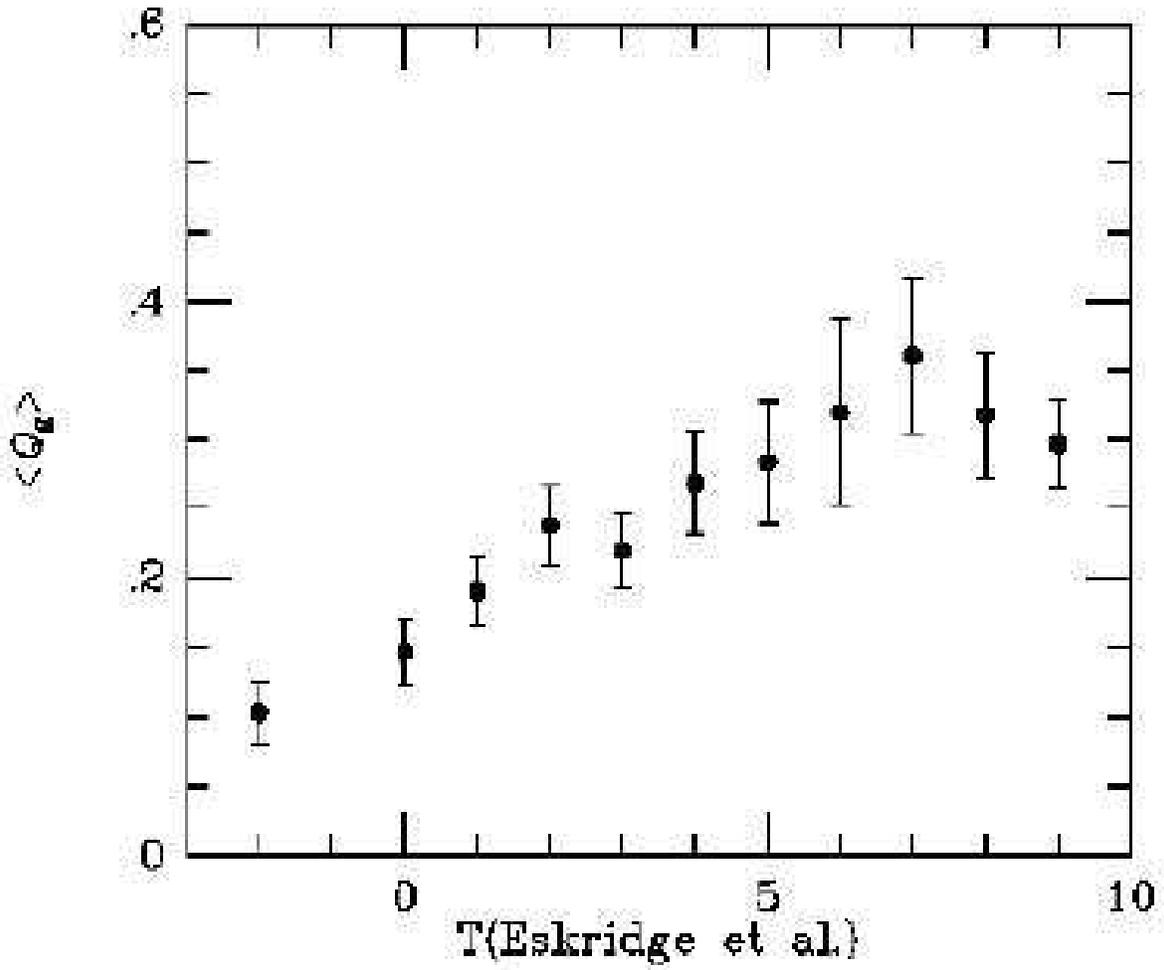}
\caption{
Plot of the mean maximum relative torque versus the near-infrared
type from Eskridge et al. (2002) for 146 OSUBGS galaxies.}
\label{esbytype}
\end{figure}

Figure~\ref{esbytype} shows that when $<Q_g>$ is
plotted against the numerically coded near-IR stages,
a strong trend with type is seen that extends
into the near-IR S0 class. The trend is smoother
than that found using RC3 types, but has about the
same amplitude from S0/a to Sm. The improved
correlation is probably not unexpected since the
appearance of the spiral arms helped to determine
the near-IR type, and the strength of the arms
can impact $Q_g$. For example, the spiral arms
in some of the OSU galaxies is virtually invisible
in the near-IR, leading to a classification of S0.
However, the implication once again is that maximum
relative torques are weaker in early-type disk 
galaxies than in late-type disk galaxies.

\section{Conclusions}

We have derived an accurate distribution of maximum relative
gravitational torques in a sample of 180 OSUBGS and 2MASS
galaxies. The sample is representative of bright galaxies, but
is biased against late-type, low-luminosity barred spirals.
It is not biased against nonbarred galaxies. The distribution
is more accurate than previous studies because of the refinement
of the gravitational torque method. We have used two-dimensional
bulge/disk/bar decomposition to eliminate the impact of 
bulge deprojection stretch on the calculated torques, and to 
derive reliable radial scalelengths that can be scaled to 
vertical scaleheights using the type dependence of $h_R/h_z$
derived by de Grijs (1998). We have also used orientation
parameters based on isophotal ellipse fits to the blue-light
images in the OSUBGS, which will be an improvement over previously 
published values for many of the galaxies.  With these refinements, we find
a higher relative frequency of low maximum relative torque galaxies compared
to Block et al. (2002). The implications for the amount
of accreted matter advocated by Block et al. (2002) remain
to be evaluated, but we expect that the revised distribution
will favor less accretion once the models 
account for the same refinements the observations have
accounted for. This will be addressed in a future paper.

We have discussed in detail the uncertainties and biases
in our distribution of gravitational torques. Because the
sample emphasizes high-luminosity systems, 
corrections for dark matter appear to be small. In the
future, further improvements could be made by obtaining
two-dimensional velocity fields of the galaxies in question.
This would facilitate the derivation of kinematic orientation
parameters, and improved deprojections.

We find a significant dependence of the mean maximum gravitational
torque on revised Hubble type. The effect persists even when
the sample is divided into bar-dominated and spiral-dominated
subsamples, and when near-infrared types from Eskridge et al.
(2002) are used in place of RC3 types. Both bars and spirals tend
to have weaker average relative torques in early-type spirals
compared to late-type spirals. The likely cause of this
is torque dilution due to the stronger bulges in 
early-type spirals. Dark matter has only a marginal
impact on this effect.

We thank the referee, F. Combes, for valuable comments
on our paper and for sending a file with her estimates
of $Q_g$ for the OSU sample. We also thank L. Whyte for sending
her table of $f_{bar}$ values.
RB acknowledges the support of NSF Grant AST-0205143 to the
University of Alabama. EL and HS acknowledge the support of
the Academy of Finland, and EL also from the Magnus Ehrnrooth
Foundation. Funding for the OSU Bright Galaxy
Survey was provided by grants from the National Science
Foundation (grants AST-9217716 and AST-9617006), with 
additional funding from the Ohio State University.
This publication also utilized images from the Two Micron
All-Sky Survey, which is a joint project of the
University of Massachusetts and the Infrared Processing
and Analysis Center of the California Institute of
Technology, funded by the National Aeronautics and
Space Administration and the National Science Foundation.
This research has also made use of the NASA/IPAC Extragalactic
Database (NED) which is operated by the Jet Propulsion 
Laboratory, California Institute of Technology, under
contract with the National Aeronautics and Space Administration.

\clearpage

\centerline{REFERENCES}

\noindent
Abraham, R. G. \& Merrifield, M. R. 2000, \aj, 120, 2835

\noindent
Bell, E. F. \& de Jong, R. S. 2001, \apj, 550, 212

\noindent
Bournaud, F. \& Combes, F. 2002, \aap, 392, 83

\noindent
Buta, R. \& Block, D. L. 2001, \apj, 550, 243 

\noindent
Buta, R., Block, D. L., \& Knapen, J. H. 2003, \aj, in press

\noindent
Buta, R. \& Combes, F. 1996, Fund. Cos. Phys., 17, 95

\noindent
Block, D. L., Puerari, I., Knapen, J. H., Elmegreen, B. G., Buta, R.,
Stedman, S., \& Elmegreen, D. M. 2001, \aap, 375, 761

\noindent
Block, D. L., Bournaud, F., Combes, F., Puerari, I., \& Buta, R. 2002,
\aap, 394, L35

\noindent
Combes, F. \& Sanders, R. H. 1981, \aap, 96, 164

\noindent
Das, M., Teuben, P. J., Vogel, S. N., Regan, M. W., Sheth, K.,
Harris, A. I., \& Jefferys, W. H. 2003, \apj, 582, 190

\noindent
de Grijs, R. 1998, \mnras, 299, 595

\noindent
de Vaucouleurs, G. 1959, Handbuch der Physik, 53, 275

\noindent de Vaucouleurs, G. et al. 1991, Third Reference Catalog
of Bright Galaxies (New York: Springer) (RC3)

\noindent
Elmegreen, B. G. \& Elmegreen, D. M. 1985, \apj, 288, 438

\noindent
Eskridge, P., Frogel, J. A., Pogge, R. W., et al. 2000, \aj, 119, 536

\noindent
Eskridge, P., Frogel, J. A., Pogge, R. W., et al. 2002, \apjs, 143, 73

\noindent
Gilmore, G. \& Reid, N. 1983, \mnras, 202, 1025

\noindent Hohl, F. 1971, ApJ, 168, 343

\noindent
Kormendy, J. \& Norman, C. 1979, \apj, 233, 539

\noindent Kormendy, J. 1993, in Galactic Bulges, IAU Symp. No. 153, H. 
DeJonghe and H. J. Habing, eds., Kluwer, Dordrecht, p. 209

\noindent
Kranz, T., Slyz, A., \& Rix, H.-W. 2003, \apj, 586, 143

\noindent
Laurikainen, E., Salo, H., \& Rautiainen, P. 2002, \mnras, 331, 880

\noindent
Laurikainen, E. \& Salo, H. 2002, \mnras, 337, 1118

\noindent
Laurikainen, E., Salo, H., Buta, R., \& Vasylyev, S. 2003, in preparation

\noindent
Lynden-Bell, D. 1979, \mnras, 187, 101

\noindent Martin, P. \& Roy, J.-R. 1994, \apj, 424, 599

\noindent McCall, M. L. 1986, \pasp, 98, 992

\noindent Miller, R. H., Prendergast, K. H., \& Quirk, W. J. 1970,
\apj, 161, 903

\noindent
Miwa, T. \& Noguchi, M. 1998, \apj, 499, 149

\noindent
Mollenh\"off, C. \& Heidt, J. 2001, \aap, 368, 16

\noindent
Noguchi, M. 1996, \apj, 469, 605

\noindent
Norman, C. A., Sellwood, J. A., \& Hasan, H. 1996, \apj, 462, 114

\noindent
Persic, M., Salucci, P., \& Stel, F. 1996, \mnras, 281, 27

\noindent
Quillen, A. C., Frogel, J. A., \& Gonz\'alez, R. A. 1994, \apj, 437, 162 (QFG)

\noindent
Roberts, W. W., Roberts, M. S., \& Shu, F. H. 1975, \apj, 196, 381

\noindent
Salo, H., Laurikainen, E., \& Buta, R. 2003, in preparation

\noindent
Sanders, R. H. \& Tubbs, A. D. 1980, \apj, 235, 803

\noindent
Sandage, A. \& Bedke, J. 1994, Carnegie Atlas of Galaxies,
Carnegie Inst. of Wash. Publ. No. 638

\noindent
Schwarz, M. P. 1981, \apj, 247, 77

\noindent
Sellwood, J. A. 2000, in Dynamics of Galaxies: From the Early Universe
to the Present, F. Combes, G. A. Mamon, \& V. Charmandaris, eds.,
San Francisco, ASP Conf. Ser. 197, p. 3.

\noindent
Sellwood, J. A. \& Moore, E. M. 1999, \apj, 510, 125

\noindent Sellwood, J. A. and Wilkinson, A. 1993, Rep. Prog. Phys. 56, 173

\noindent
S\'ersic, J. L. 1968, Atlas de Galaxias Australes (Cordoba: Obs. Astron. Univ.
Nac. Cordoba)

\noindent Skrutskie, M. F. et al. 1997, in The Impact of Large-Scale Near-IR
Surveys, F. Grazon et al., eds., Dordrecht, Kluwer, p. 25

\noindent
Tully, R. B. 1988, Nearby Galaxies Catalogue, Cambridge, Cambridge
University Press

\noindent
Whyte, L., Abraham, R. G., Merrifield, M. R., Eskridge, P. B., Frogel,
J. A., \& Pogge, R. W. 2002, \mnras, 336, 1281

\clearpage

\begin{deluxetable}{lrrrrrr}
\tablewidth{0pc}
\tablecaption{Sample Properties\tablenotemark{a}}
\tablehead{
\colhead{Parameter} &
\colhead{SA} & 
\colhead{SAB} &
\colhead{SB} &
\colhead{SA (T88)} & 
\colhead{SAB (T88)} &
\colhead{SB (T88)}
}
\startdata
$n$ & 58 & 57 & 62 & 291 & 364 & 609 \\
$<A_B(G)>$ & 0.15 & 0.10 & 0.17 & 0.19 & 0.17 & 0.21 \\
$<logR_{25}>$ & 0.17 & 0.15 & 0.15 & 0.29 & 0.21 & 0.30 \\
$<T>$  & 3.67 & 3.83 & 3.61 & 4.45 & 5.18 & 6.46 \\
$<logD_o>$ & 1.64 & 1.66 & 1.62 & 1.58 & 1.51 & 1.46 \\
$<(B-V)_T^o>$ & 0.62 (51) & 0.61 (49) & 0.60 (52) & 0.58 (209) & 0.56 (219) & 0.53 (286) \\
$<(U-B)_T^o>$ & 0.06 (41) & 0.05 (39) & 0.04 (46) & 0.00 (169) & $-$0.03 (175) & $-$0.07 (256) \\
$<V_{\odot}>$ (km s$^{-1}$) & 1467 & 1322 & 1536 & 1564 & 1622 & 1543 \\
$<\Delta>$ (Mpc) & 21.0 & 19.0 & 20.8 & 22.3 & 23.6 & 21.6 \\
$<M_B^o>$ & $-$20.37 & $-$20.21 & $-$20.22 & $-$19.8 & $-$19.5 & $-$18.7 \\
$<D_o>$ (kpc) & 26.7 & 25.5 & 25.4 & 24.3 & 22.9 & 18.5 \\
$<Q_g>\pm\sigma$ & 0.110$\pm$0.065 & 0.221$\pm$0.122 & 0.331$\pm$0.147 
& ..... & ..... & ..... \\
\enddata
\tablenotetext{a}{Numbers in parentheses are the sample sizes available for
the indicated mean parameters. T88 refers to the catalogue of Tully (1988).}
\end{deluxetable}

\begin{deluxetable}{crc}
\tablewidth{0pc}
\tablecaption{Distribution of Maximum Relative Torques}
\tablehead{
\colhead{$Q_g$} &
\colhead{$n$} & 
\colhead{$f$} 
}
\startdata
     0.025   &     10  &   0.056\\
     0.075   &     32  &   0.178\\
     0.125   &     29  &   0.161\\
     0.175   &     27  &   0.150\\
     0.225   &     17  &   0.094\\
     0.275   &     16  &   0.089\\
     0.325   &     14  &   0.078\\
     0.375   &     12  &   0.067\\
     0.425   &     10  &   0.056\\
     0.475   &      2  &   0.011\\
     0.525   &      6  &   0.033\\
     0.575   &      0  &   0.000\\
     0.625   &      2  &   0.011\\
     0.675   &      3  &   0.017\\
\enddata
\end{deluxetable}

\begin{deluxetable}{cccccc}
\tablewidth{0pc}
\tablecaption{Uncertainties}
\tablehead{
\colhead{$<i>$} &
\colhead{$<Q_g>$} & 
\colhead{ave. dev.(i$\pm$5)} &
\colhead{ave. dev.(pa$\pm$4)} &
\colhead{ave. dev.($h_z$)} &
\colhead{$n$} 
}
\startdata
    24.0  &   0.237  &   0.010   & 0.009 & 0.032 &   39 \\
    40.5  &   0.237  &   0.019   & 0.010 & 0.027 &   83 \\
    58.5  &   0.190  &   0.038   & 0.020 & 0.021 &   58 \\
\enddata
\end{deluxetable}

\begin{deluxetable}{lrcccc}
\tablewidth{0pc}
\tablecaption{Inclination Effects}
\tablehead{
\colhead{Sample} &
\colhead{$n$} &
\colhead{$<Q_g>\pm m.e.$} & 
\colhead{$\sigma$} &
\colhead{$<Q_g>\pm m.e.$} & 
\colhead{$\sigma$} \\
\colhead{} &
\colhead{} &
\colhead{$i \leq i_{med}$} & 
\colhead{} &
\colhead{$i \geq i_{med}$} & 
\colhead{} \\
}
\startdata
SA   &  58 & 0.115$\pm$0.015 & 0.080 & 0.104$\pm$0.009 & 0.046 \\
SAB  &  57 & 0.245$\pm$0.026 & 0.141 & 0.196$\pm$0.018 & 0.094 \\
SB   &  62 & 0.325$\pm$0.027 & 0.148 & 0.336$\pm$0.027 & 0.148 \\
full & 180 & 0.233$\pm$0.016 & 0.151 & 0.211$\pm$0.015 & 0.143 \\
\enddata
\end{deluxetable}

\begin{deluxetable}{lccccr}
\tablewidth{0pc}
\tablecaption{Mean Maximum Relative Torque by Optical Revised Hubble Type}
\tablehead{
\colhead{Stage} &
\colhead{$T$(RC3)} &
\colhead{$<Q_g>$} &
\colhead{$\sigma$} & 
\colhead{mean error} &
\colhead{$n$} 
}
\startdata
S0/a &         0  &   0.195  &   0.131  &   0.038   &     12 \\
Sa   &         1  &   0.125  &   0.108  &   0.028   &     15 \\
Sab  &         2  &   0.155  &   0.124  &   0.030   &     17 \\
Sb   &         3  &   0.205  &   0.129  &   0.023   &     32 \\
Sbc  &         4  &   0.242  &   0.140  &   0.022   &     39 \\
Sc   &         5  &   0.246  &   0.155  &   0.025   &     38 \\
Scd  &         6  &   0.321  &   0.180  &   0.050   &     13 \\
Sd   &         7  &   0.224  &   0.137  &   0.056   &      6 \\
Sdm  &         8  &   0.331  &   0.258  &   0.149   &      3 \\
Sm   &         9  &   0.328  &   0.066  &   0.038   &      3 \\
     &            &          &          &           &        \\
S0/a-Sb &     0-3 &   0.177  &   0.126  &   0.014   &     76 \\
Sbc-Sm  &     4-9 &   0.258  &   0.153  &   0.015   &    102 \\
     &            &          &          &           &        \\
SA0/a-SAb &     0-3 &   0.068  &   0.038  &   0.008   &     24 \\
SAbc-SAm  &     4-9 &   0.139  &   0.064  &   0.011   &     34 \\
     &            &          &          &           &        \\
SAB0/a-SABb &     0-3 &   0.145  &   0.073  &   0.017   &     19 \\
SABbc-SABm  &     4-9 &   0.260  &   0.124  &   0.020   &     38 \\
     &            &          &          &           &        \\
SB0/a-SBb &     0-3 &   0.274  &   0.118  &   0.021   &     33 \\
SBbc-SBm  &     4-9 &   0.395  &   0.152  &   0.028   &     29 \\
\enddata
\end{deluxetable}

\begin{deluxetable}{lcccr}
\tablewidth{0pc}
\tablecaption{Mean Maximum Relative Torque by Near-Infrared Revised Hubble Classification\tablenotemark{a}}
\tablehead{
\colhead{Stage or Family} &
\colhead{$<Q_g>$} &
\colhead{$\sigma$} & 
\colhead{mean error} &
\colhead{$n$} 
}
\startdata
S0    &        0.103  &   0.070  &   0.022   &     10\\
S0/a  &        0.147  &   0.095  &   0.024   &     15\\
Sa    &        0.191  &   0.124  &   0.025   &     24\\
Sab   &        0.238  &   0.121  &   0.029   &     18\\
Sb    &        0.220  &   0.143  &   0.027   &     28\\
Sbc   &        0.269  &   0.168  &   0.037   &     20\\
Sc    &        0.284  &   0.152  &   0.044   &     12\\
Scd   &        0.320  &   0.200  &   0.067   &      9\\
Sd    &        0.361  &   0.177  &   0.056   &     10\\
Sdm   &        0.318  &   0.111  &   0.045   &      6\\
Sm    &        0.297  &   0.063  &   0.032   &      4\\
      &               &          &           &       \\
S0-Sb  &        0.159 &   0.110  &   0.016   &     49\\
Sbc-Sm &        0.265 &   0.158  &   0.016   &     97\\
      &               &          &           &       \\
      &               &          &           &       \\
S     &         0.116 &   0.082  &   0.014   &     32\\
SAB   &         0.174 &   0.112  &   0.022   &     26\\
SB    &         0.290 &   0.147  &   0.015   &     98\\
\enddata
\tablenotetext{a}{Classifications are from Col. 5 of Table 1 of
Eskridge et al. (2002).}
\end{deluxetable}

\clearpage

\end{document}